\documentclass[12pt,preprint]{emulateapj}

\usepackage{amsmath}
\slugcomment{ApJ in press}

\shorttitle{The unusual X-ray morphology of NGC~4636}
\shortauthors{Baldi et al.}

\usepackage[below]{placeins}
\usepackage{natbib}
\usepackage{journal_shortcuts}
\bibpunct{(}{)}{;}{a}{}{;}

\begin{document}

\title{The unusual X-ray morphology of NGC~4636 revealed by deep Chandra observations: cavities and shocks created
by past AGN outbursts}


\author{A. Baldi, W. Forman, C. Jones, R. Kraft, P. Nulsen}
\affil{Harvard-Smithsonian Center for Astrophysics}

%\author{W. Forman}
%\affil{Harvard-Smithsonian Center for Astrophysics}
%\email{aastex-help@aas.org}

%\author{C. Jones}
%\affil{Harvard-Smithsonian Center for Astrophysics}

%\author{R. Kraft}
%\affil{Harvard-Smithsonian Center for Astrophysics}

%\author{P. Nulsen}
%\affil{Harvard-Smithsonian Center for Astrophysics}

\author{E. Churazov}
\affil{Max-Planck-Institute f\"ur Astrophysik}

\author{L. David, S.Giacintucci}
\affil{Harvard-Smithsonian Center for Astrophysics}

%\author{L. David}
%\affil{Harvard-Smithsonian Center for Astrophysics}

%\and
%
%\author{S. Giacintucci}
%\affil{Harvard-Smithsonian Center for Astrophysics}

%\author{et al.}


%\altaffiltext{1}{Visiting Astronomer, Cerro Tololo Inter-American Observatory.
%CTIO is operated by AURA, Inc.\ under contract to the National Science
%\altaffiltext{2}{Society of Fellows, Harvard University.}
%\altaffiltext{3}{present address: Center for Astrophysics,
%    60 Garden Street, Cambridge, MA 02138}
%\altaffiltext{4}{Visiting Programmer, Space Telescope Science Institute}
%\altaffiltext{5}{Patron, Alonso's Bar and Grill}


\begin{abstract}
We present Chandra ACIS-I and ACIS-S observations ($\sim$200~ks in total) of the X-ray luminous 
elliptical 
galaxy NGC 4636, located in the outskirts of the Virgo cluster.
A soft band (0.5-2 keV) image shows the presence of a bright core in the center surrounded by an 
extended 
X-ray corona and two pronounced quasi-symmetric, 8 kpc long, arm-like features. Each of this 
features defines the rim
of an ellipsoidal bubble. An additional bubble-like feature, whose northern rim is located 
$\sim2$~kpc south of the 
north-eastern arm, is detected as well.
We present surface brightness and temperature profiles across the rims of the bubbles, showing 
that their edges are sharp 
and characterized by temperature jumps of about 20-25\%. 
Through a comparison of the observed profiles with theoretical shock models, we demonstrate that 
a scenario 
where the bubbles were produced by shocks, probably driven by energy deposited off-center by jets, 
is the most 
viable explanation to the X-ray morphology observed in the central part of NGC 4636. As a 
confirmation to this
scenario, radio jets extending towards the bubbles and a central weak X-ray and radio source are 
detected and
are most likely the signs of AGN activity which was more intense in the past.
A bright dense core of $\sim$1~kpc radius is observed at the center of NGC~4636.
A sharp decline in surface brightness from the core to the ambient gas is observed and is not 
accompanied by 
a variation in the temperature and thus could not be in thermal pressure equilibrium. 
However the bright core could be a long lived feature if the radio jets
are acting as a balancing factor to thermal pressure or if the bright core is produced by steep
abundance gradients.
\end{abstract}



%% tied to the reference list via symbolic KEYs. The KEY corregalaxies: nuclei sponds


\section{Introduction}

It is well known that active galactic nuclei (AGN) play an important role in the
evolution of the hot gas in both individual galaxies and clusters of galaxies
\citep[e.g. ][]{churazov00,bruggen02,cavaliere02,hoeft04,forman05,mcnamara07}.
The so-called `AGN feedback' can cause re-heating of the central cooling regions
of a cluster and balance the cooling due to the X-ray emission \citep[see][for a review]{mcnamara07}.
AGN are also effective in shaping the morphology of the hot gas halos around individual
galaxies and clusters, e.g. giving rise to cavities and bubbles \citep{birzan04}.\\
While several examples of subsonic bubble inflation due to AGN outbursts in individual galaxies are present
in the literature \citep[e.g. NGC~507, M~84; ][]{kraft04,finoguenov08}, supersonic bubble expansion,
giving rise to shocks in the hot X-ray emitting gas, has only been observed in a handful of cases
\citep[e.g. M~87, NGC~4552; ][]{forman05,machachek06}.
Another possible case of supersonic bubble expansion could be that of NGC 4636, 
the dominant galaxy of a group on the outskirts of the Virgo Cluster 
\citep[10$^\circ$ or 2.6 Mpc on the sky to the south of M87, at a distance to NGC~4636 of 15~Mpc;][]{tonry01}.
%The optical surface brightness profile of NGC 4636 flattens in the inner regions
%as expected for a luminous, slowly rotating elliptical galaxy. As observed by Hubble Space
%Telescope exposures, the central region has low eccentricity and this galaxy is thus 
%classified as an E0, although, at low surface brightness, 
%the galaxy is flattened \citep[E4;][]{sandage61}. The H$\alpha$ luminosity is measured to be 
%$2.3\times10^{38}$~ergs s$^{-1}$ \citep{ho97}.
This galaxy has been observed extensively 
by every major X-ray imaging observatory to date. 
Spectral imaging with Einstein showed the galaxy 
to be surrounded by an extensive halo of hot gas with a temperature 
of 0.78~keV \citep{forman85}. 
Further studies using ROSAT and ASCA found gradients in both the temperature and abundance 
profiles, along with a very extended halo
\citep{awaki94,trinchieri94,matsushita97,finoguenov00,buote00,osullivan03}. 
The metallicity is about solar in the central 9~kpc ($2^\prime$) region, decreasing to 
0.2-0.3 solar at $\sim45$~kpc ($10^\prime$) from the galaxy center \citep{matsushita97}.
%Although the galaxy is located in the outskirts of the Virgo Cluster, X-ray 
%emission associated with Virgo has not been observed around NGC 4636.
More recently, NGC~4636 has been 
observed by both Chandra and XMM-Newton. The initial Chandra ACIS-S 
observation showed an unusual morphology in the core of the galaxy, 
most notably quasi-symmetric spiral arm-like features and a possible ring of 
high surface brightness emission, thought to be the product of shocks 
driven by a previous AGN outburst 
\citep{jones02}.
As a confirmation to the outburst driven shock scenario, a change 
of the gas temperature across one of the arms was 
observed as well. 
From the same Chandra data, \citet{ohto03} performed a detailed spectral 
analysis, suggesting 
the presence of a cavity in the X-ray halo on the west side of the core, with
AGN activity being the most likely cause also in this case. In this scenario, plasma from 
the AGN jets may be able to displace the gas of the X-ray halo, leaving 
an apparent void as currently observed in the X-ray morphology.
Using data from XMM-Newton and Chandra, \citet{osullivan05} mapped the 
temperature and abundance structure of the inner halo of NGC 4636, confirming the 
detection of a cavity to the west of the galaxy and revealing the presence of 
a plume of cool metal-rich gas extending beyond the cavity to the southwest. 
Both cavity and plume appear to be the product of past AGN activity. Moreover, their
abundance determination shows a depression in the central few kpc ($\sim0.2$ solar)
in disagreement with the results of \citet{matsushita97}.   
%An upper limit of $2.7\times10^{38}$~ergs s$^{-1}$ is 
%estimated for the present X-ray emission of the AGN \citep{loewenstein01}.\\

In this paper we take advantage of the long Chandra ACIS-I
observations ($\sim150$~ks) combined with the previously observed ACIS-S 
exposure ($\sim$40~ks) to analyze in more detail the features revealed by the previous
Chandra and XMM-Newton observations. 
In particular we are able to study in more detail the morphology and
temperature structure in the observed cavities and look for gradients in temperature
and abundance to test the shock scenario invoked for their formation
\citep{jones02,ohto03,osullivan05}. Moreover we will investigate with unprecedented detail 
the morphology and the properties of the hot gas in the galaxy core.

At the distance of NGC~4636 \citep[15~Mpc;][]{tonry01},
$1^{\prime\prime}\approx70$~pc. Throughout the paper, 
all the uncertainties are quoted at 1$\sigma$ (68\%) for one
interesting parameter, unless otherwise stated.
The abundance estimates are relative to the
cosmic values given in \citet{anders89}.

\section{Chandra Data Preparation and Analysis} \label{dataprep}

NGC~4636 was observed twice by ACIS-S and twice by ACIS-I.
In our analysis, we use the longer ACIS-S observation (performed on 2000, January 26; ObsID: 323) 
and both the ACIS-I observations, performed one immediately after the other 
on 2003, February 14 (ObsID: 3926) and on 2003, February 15 (ObsiD: 4415).
The Chandra data analysis has been performed using CIAO 
v3.4, which applies the newest ACIS gain maps, time-dependent ACIS gain corrections, 
and ACIS charge transfer inefficiency corrections. 
The background light curve during the observation has been examined to 
detect periods of high background 
following \citet{markevitch03}. A flaring episod (lasting $\sim15$~ks) is
observed during the ACIS-S observation, reducing the usable exposure time to $\sim37$~ks.
No significant
flaring episodes are detected during the two ACIS-I observations, giving a total good 
integration time for the ACIS-I detector of $\sim149$~ks.\\

\subsection{Background Subtraction}\label{bkgsub}
%Since the hot medium around NGC~4636 fills most of the Chandra field of view, even in ACIS-I, 
We use the ``blank-field'' observations, processed identically to the galaxy observations 
(i.e. as described above) and reprojected onto the sky using
the aspect information from the galaxy pointings. 
%It is worth noticing that the 
%synthetic backgrounds correspond by far to a longer exposure time ($\sim0.5$ Msec) 
%than our observation, giving us a good estimate of the 
%background.
%Moreover, 
%\footnote{$http://cxc.harvard.edu/cal/Acis/Cal\_prods/bkgrnd/acisbg/COOKBOOK$}. 
For the ACIS-I observations, the ``blank-fields'' are renormalized  
to the background of the observation using the ACIS-S2 chip.
%, in a region of the ACIS field 
%of view practically free from galaxy emission.
For the ACIS-S observation, the chip used for the renormalization is ACIS-S1 owing
to the extended source emission on S3. 
%Since the observation was performed
%using ACIS-I, we selected the ACIS-S2 chip as the most suitable for our purpose.
The energy band from 9.5 to 12 keV (mostly
dominated by charged particles) is used to perform the normalization.\\
%and if appropriate adjustments are needed. 
We follow the procedure of \citet{vikhlinin05}, to check
%In addition to the particle-induced background we checked also 
if the diffuse soft X-ray background could be an important background component
in our observation. A 
spectrum from the source-free region of the S2 chip was extracted, 
the renormalized ``blank-field'' background was subtracted and the 
background subtracted spectrum was fitted in
XSPEC v11.3.2p (in the 0.4-1 keV band) with an unabsorbed {\tt mekal}
model, whose normalization is allowed to be negative.
The residuals are found to be negligible and consistent with zero in all three observations, 
suggesting that no additional soft background correction is necessary for our datasets.

\section{Results}

\subsection{X-ray morphology and surface brightness}\label{radanal}
\begin{figure}
%\epsscale{.80}
\plottwo{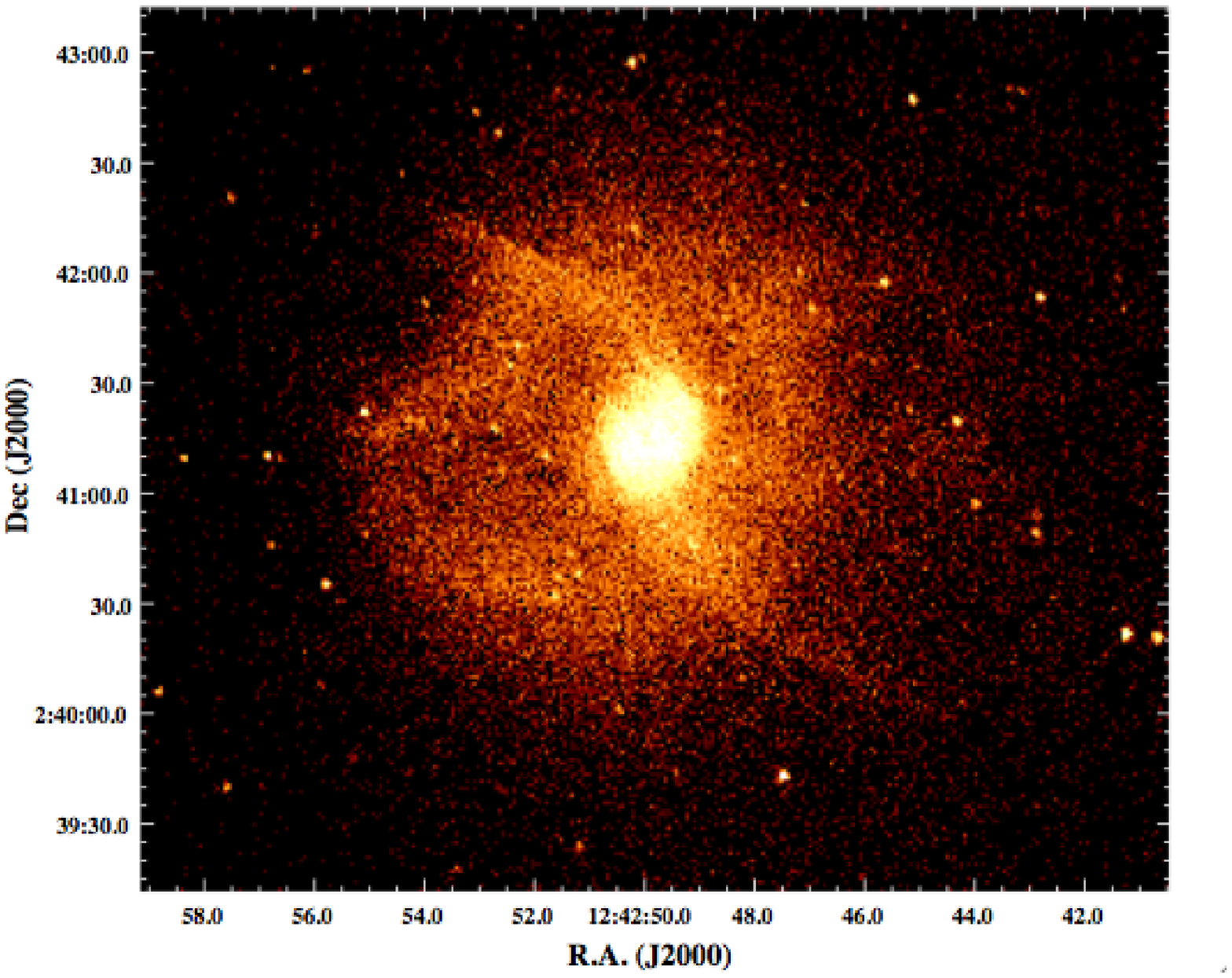}{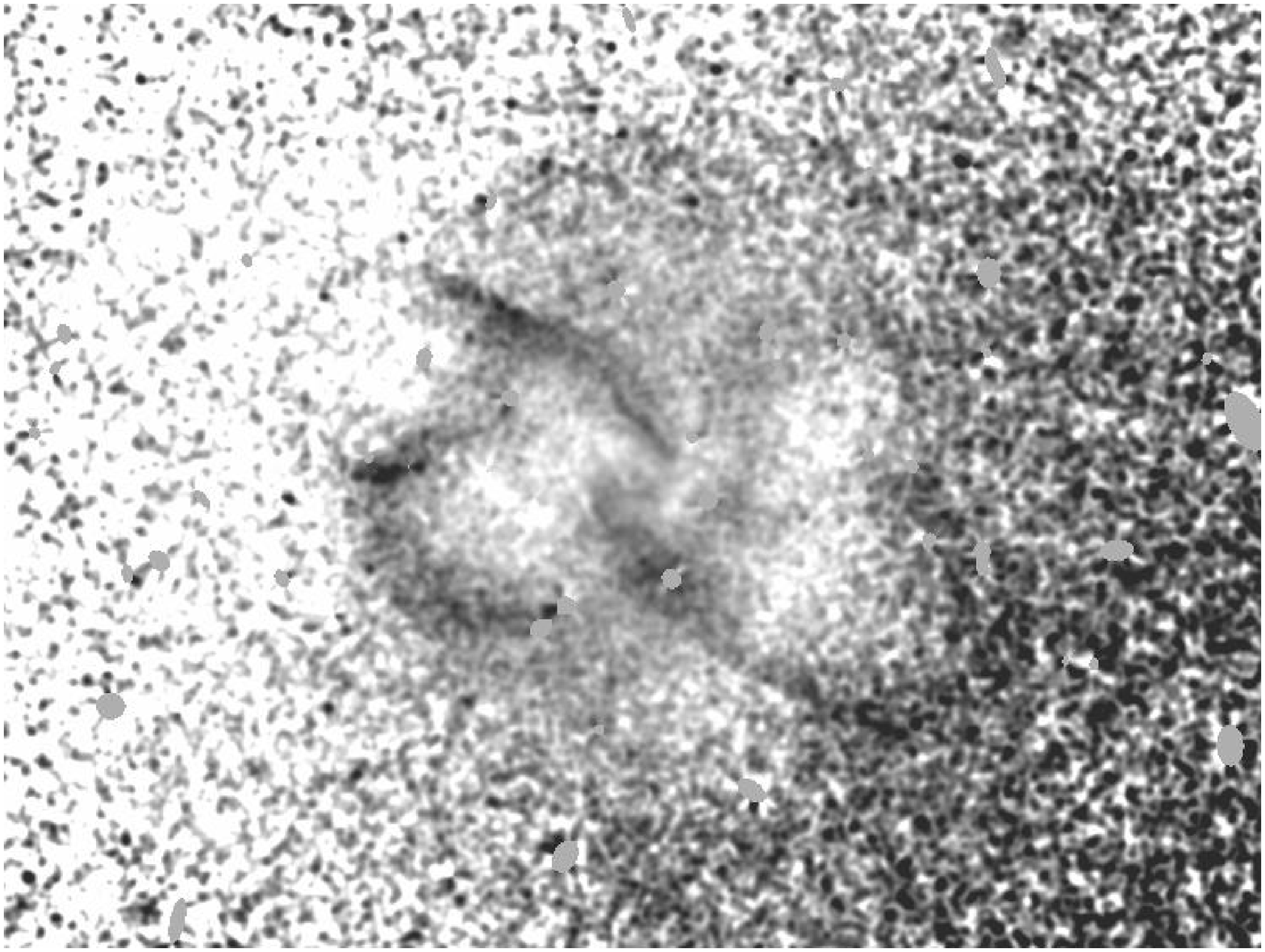}
\caption{{\it Left:} Chandra ACIS-I+ACIS-S image of NGC~4636 in the 0.3-2 keV band. The galaxy shows the presence
of a bright central core ($r\sim1~kpc\equiv15^{\prime\prime}$), surrounded by a lower surface brightness region 
extending as far as $\sim6$~kpc ($80^{\prime\prime}$) from the center.
Two pronounced quasi-symmetric (8~kpc long) arm-like features define the rims of ellipsoidal bubbles. 
Also an additional bubble-like feature, whose northern rim is located $\sim2$~kpc south of the 
north-eastern arm, is visible from the Chandra image.
{\it Right:} Chandra ACIS-I+ACIS-S image after the subtraction of a $\beta$-model fitted to
the general diffuse X-ray emission. The rims of the ellipsoidal bubbles are more clearly
visible and another bubble almost symmetric to the bubble-like feature observed south of the
north-eastern arm, could be present north-west of the nucleus.
\label{xrayimage}}
\end{figure}

\begin{figure}
%\epsscale{.80}
\plotone{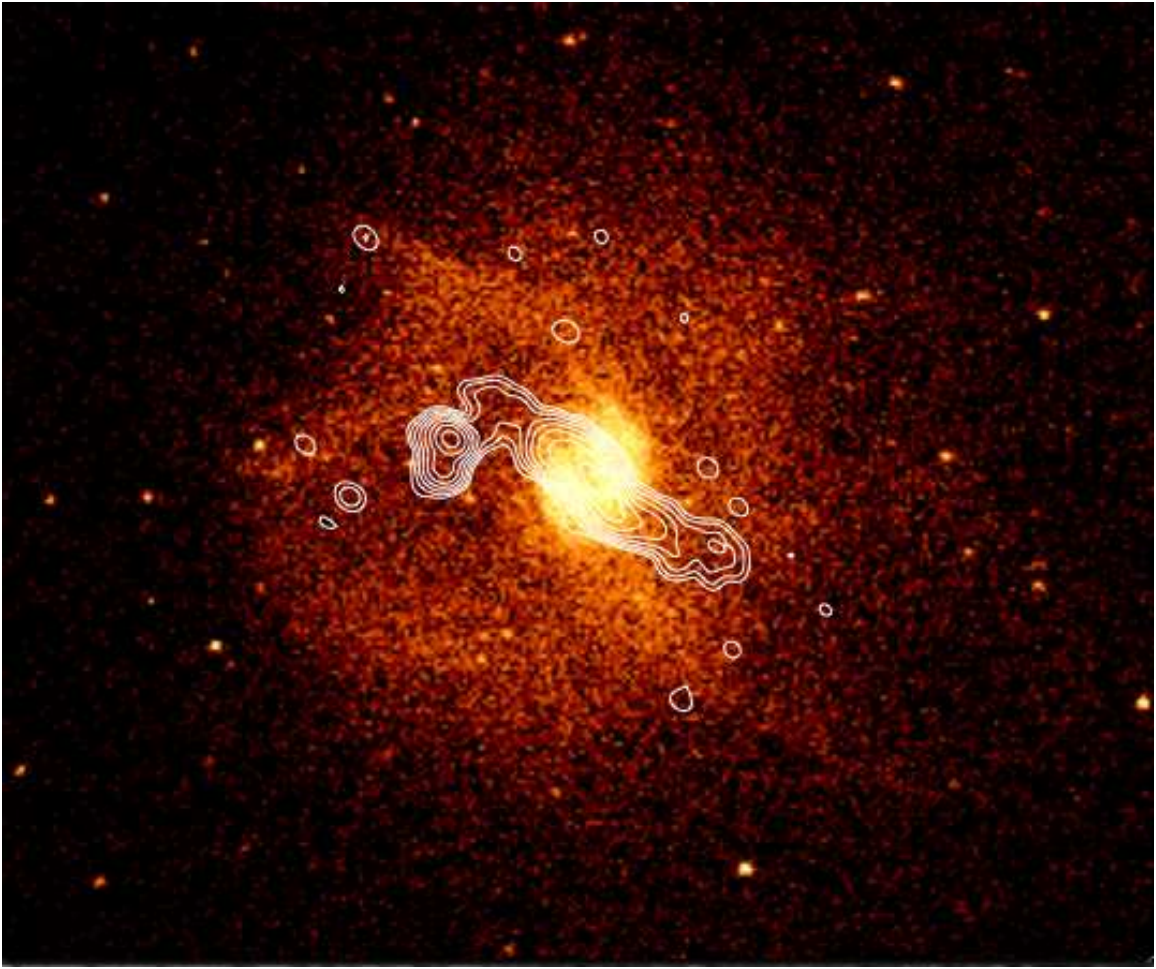}
\caption{Chandra ACIS-I+ACIS-S image (0.5-3 keV) of NGC~4636 with the 610~MHz GMRT radio contours
superimposed (from Giacintucci et al. in prep.). 
The radio contours shows a correlation with the cavities, however they 
do not fill them completely, or break through the cavity boundaries. The resolution of the
radio image is $5.8^{\prime\prime}\times 4.3^{\prime\prime}$, p.a. $48^\circ$, with the contours 
spaced by a factor 2 starting from $\pm3\sigma=0.15$ mJy/beam.
An unrelated radio point-source is visible on the left of the nucleus, just below the NE cavity.
\label{xrayradio}}
\end{figure}

Fig.~\ref{xrayimage}a shows the merged 0.3-2~keV Chandra image of all three observations (two ACIS-I 
and one ACIS-S observation).
The image was ``flat-fielded'' using an appropriate exposure map weighted for the emission of a 0.6 keV
thermal plasma.
The galaxy presents a very bright central core of radius $\sim1~kpc$ (15$^{\prime\prime}$) centered on
the nucleus.
A lower surface brightness region surrounds the nucleus extending out to $\sim6$~kpc ($80^{\prime\prime}$).
Two quasi-symmetric arm-like features are embedded in this lower surface brightness emission. These features were 
previously reported by \citet{jones02} who analyzed the shorter ACIS-S observation. However, combining
these data with the two ACIS-I observations we are able to observe that the south-western arm is clearly
part of an X-ray cavity extending as far as $\sim9$~kpc ($2^{\prime}$) from the center. 
The hint of a similar structure can 
be observed in coincidence with the north-eastern arm-like feature. It is not clear from the X-ray
image whether there is a cavity in this case.
However the cavities are more evident if we remove the contribution from the general diffuse
emission of the galaxy. Fig.~\ref{xrayimage}b shows the merged Chandra image in the 0.3-2 keV band where
a $\beta$-model fitted to the galaxy diffuse emission was subtracted. This processing allowed to highlight
fainter structures.
Indeed, both the south-western and the north-eastern cavities are clearly visible in this Figure.
Moreover two additional quasi-symmetric cavities can be seen, just to the East of the nucleus and to
the North-West of it, with the rims of the former being better defined than the latter.\\ 
The radio data overlaid on the X-ray image (Fig.~\ref{xrayradio}) shows a correlation with the cavities.
However, the radio lobes do not fill the SW cavity, or break through the cavity boundaries
as in other galaxies presenting cavities \citep[e.g. M~84,][]{finoguenov08}. However, the lobes
detected at 610~MHz (Giacintucci et al. in prep.) with the Giant Metrewave Radio Telescope (GMRT) extend 
further than those 
observed at 1.4~GHz and also the NE cavity seems to be
filled by plasma detected at 235~MHz with the GMRT (Giacintucci et al. in prep.). The
radio lobes are weak, having a luminosity at 1.4~GHz of $\sim1.4\times10^{38}$ erg s$^{-1}$.
However the plasma creating
the cavities remains partially undetected by the radio observations of NGC~4636 performed so far.

\begin{figure}
%\epsscale{.80}
\plotone{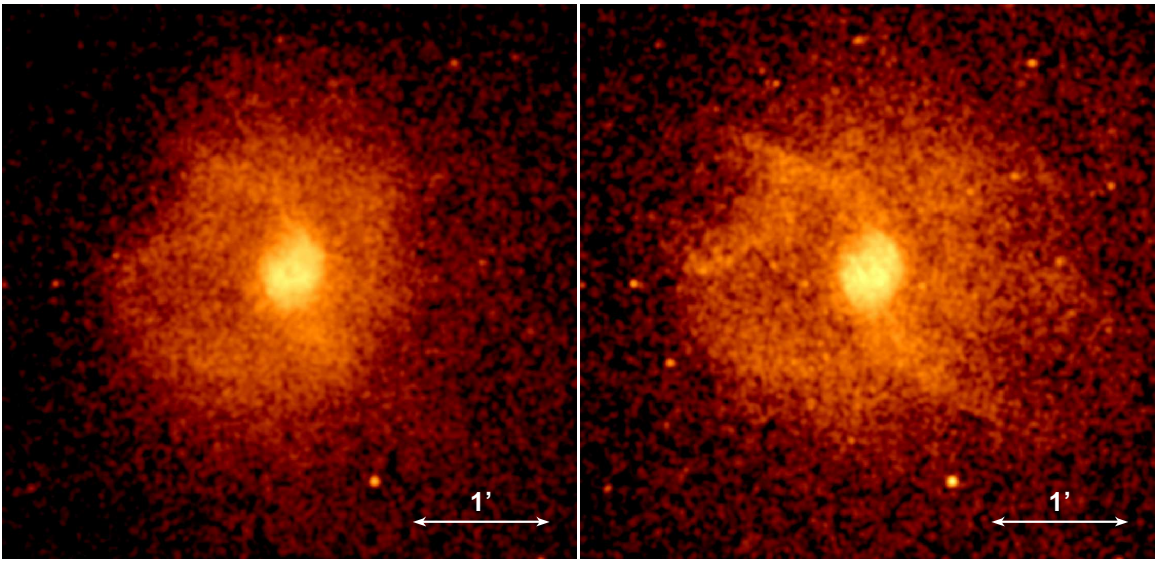}
\caption{{\it Left:} 0.3-0.9 keV Chandra ACIS-I+ACIS-S image of NGC~4636 smoothed with a
$2^{\prime\prime}$ width gaussian; {\it Right:}
0.9-1.3 keV Chandra ACIS-I+ACIS-S image of NGC~4636 smoothed with a
$2^{\prime\prime}$ width gaussian. The X-ray arm-like structures
become prominent only in the higher energy band. In the figure $1^{\prime\prime}$
corresponds to $\sim4.2$~kpc. \label{hardsoft}}
\end{figure}

Figure~\ref{hardsoft} shows Chandra images of NGC~4636 in two different bands: 0.3-0.9 keV and 0.9-1.3 keV.
A $2^{\prime\prime}$ width gaussian smoothing was applied to them to enhance the features visible in the two
images.
The morphology is quite different between them, with the X-ray arm-like features becoming prominent only at 
energies larger then 0.9 keV, a possible indication that the temperature in these features is higher than
in the surrounding hot interstellar medium of the galaxy.

\subsection{Global X-ray properties}\label{global}
\begin{figure}
%\epsscale{.80}
\plotone{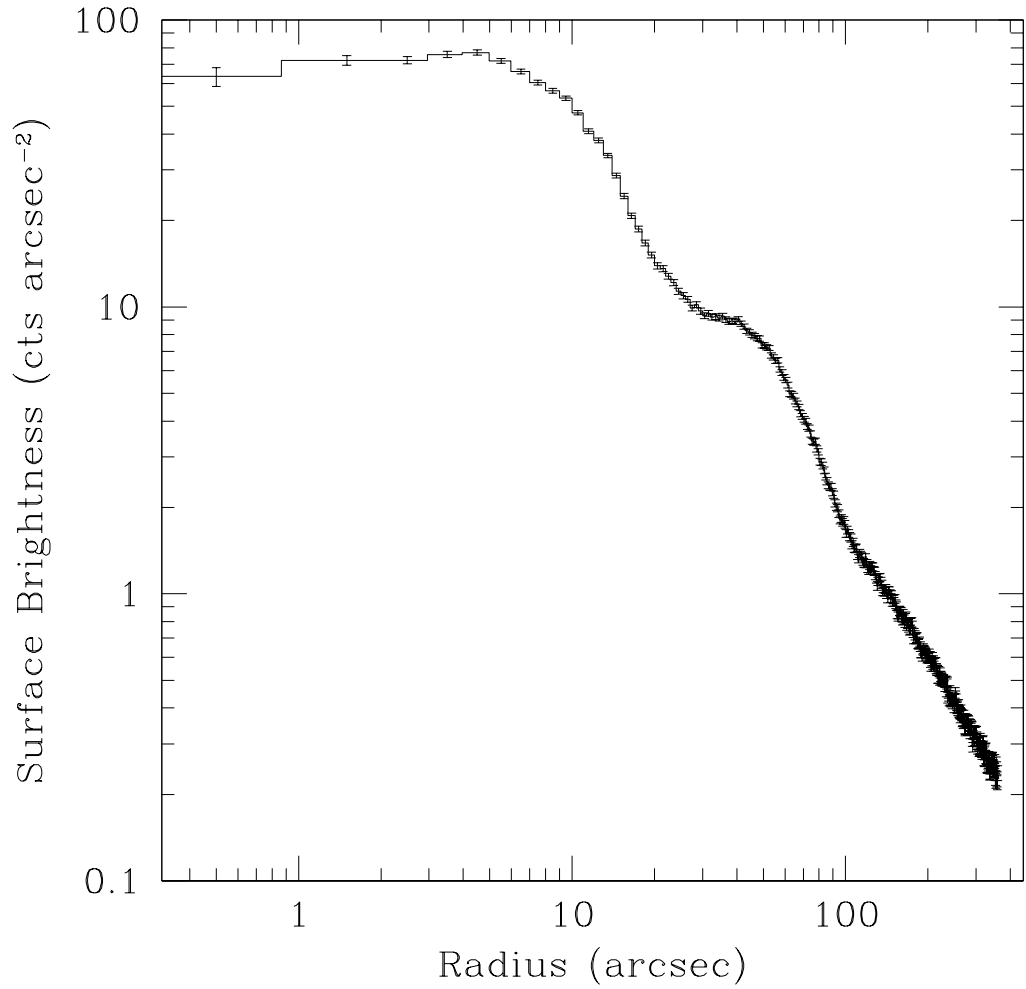}
\caption{Chandra ACIS-I+ACIS-S surface brightness profile of NGC~4636 extracted in the 0.5-3 keV band.
A central surface brightness peak is not present, being quite flat at the center. The profile declines monotonically 
at $5^{\prime\prime}<r<25^{\prime\prime}$ ($0.35$~kpc $<r<1.75$~kpc), showing another `plateau' at 
$25^{\prime\prime}<r<60^{\prime\prime}$ ($1.75$~kpc $<r<4.2$~kpc).
\label{sbprof}}
\end{figure}

\begin{figure}
%\epsscale{.80}
\plottwo{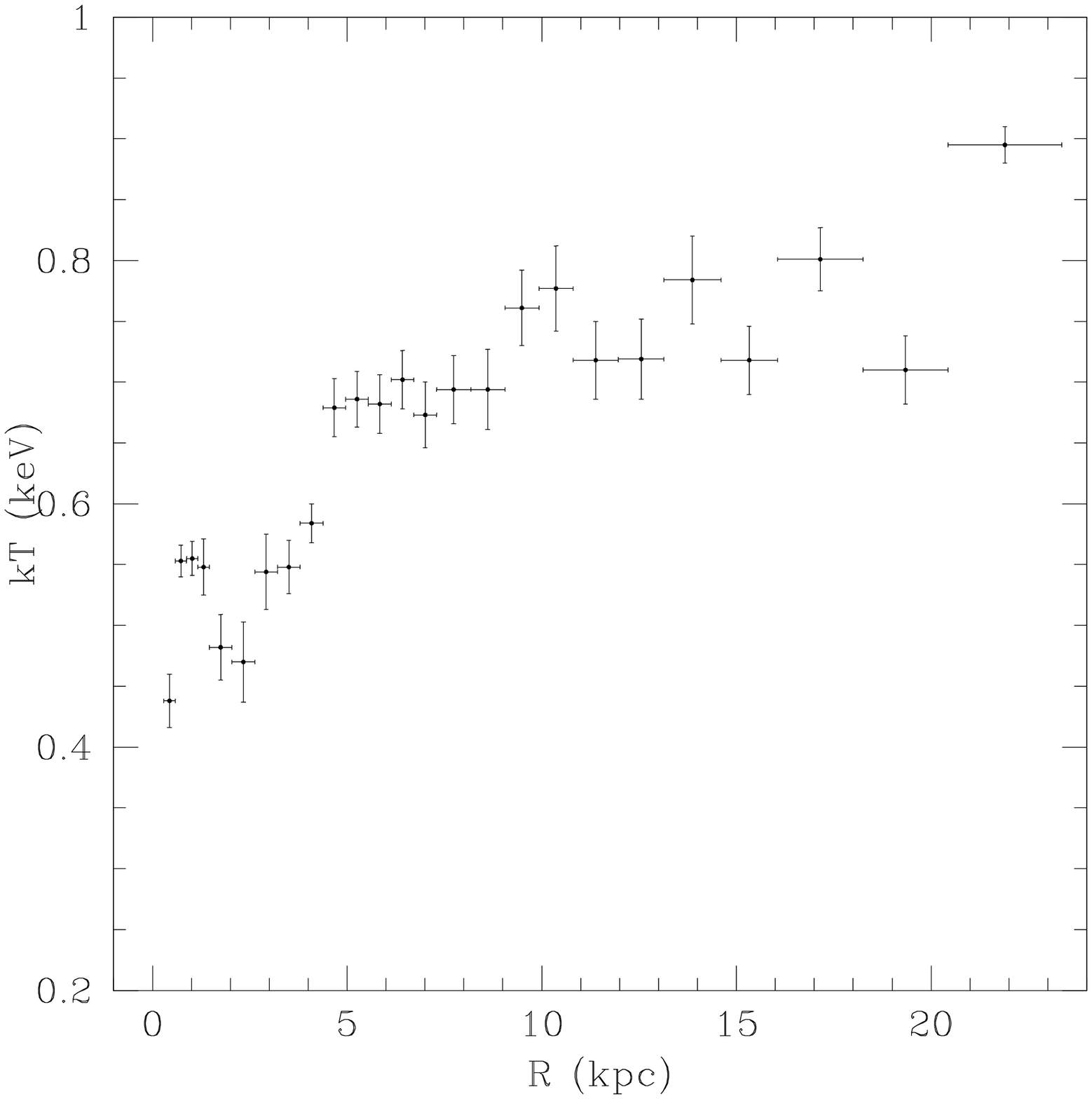}{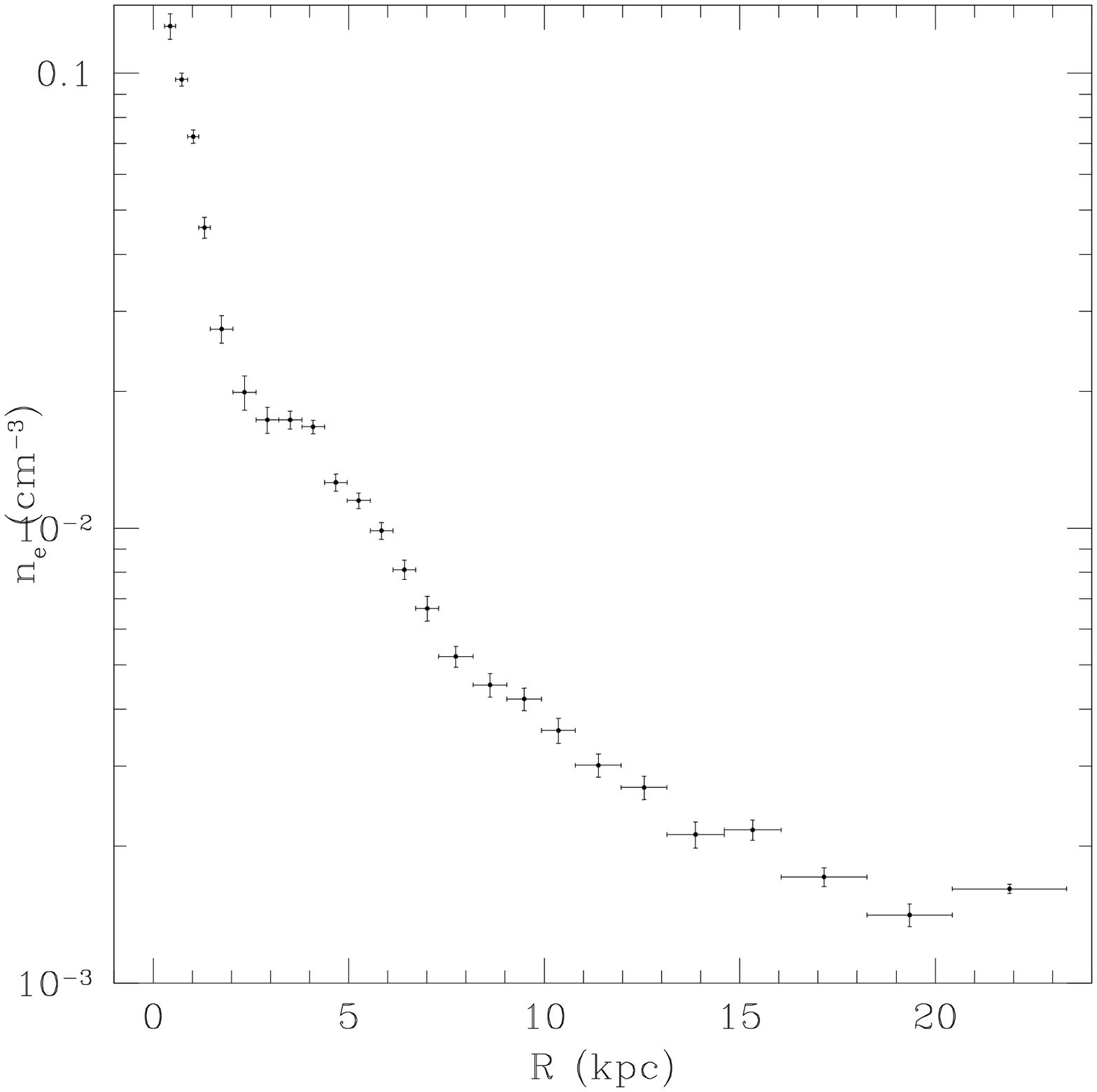}
\caption{{\it Left:} Deprojected temperature profile for the X-ray gas observed in NGC~4636. A decline of the
temperature is observed at the center, where $kT\sim0.5$~keV, while the temperature is increasing toward the 
outskirts of the galaxy ($kT\ga 0.8$~keV).{\it Right:} 
Deprojected electron density profile for NGC~4636. The density shows the presence of a central
peak and of a `plateau' coincident with the flattening observed in the surface brightness profile, located at 
2~kpc$\la r\la$4.5~kpc from the center.\label{ktz}}
\end{figure}

%Although the morphology of the X-ray emission around the nucleus of NGC~4636 is extremely complex, it could
%be useful to study its average radial properties.
The surface brightness profile of NGC~4636 is shown in Figure~\ref{sbprof}. The emission does not show a central
peak and is quite flat at the center, then it starts to decline monotonically at $\sim5^{\prime\prime}$ ($0.35$~kpc) from
the center, showing
another `plateau' at $25^{\prime\prime}<r<60^{\prime\prime}$ ($1.75$~kpc$<r<4.2$~kpc). Both these features can be noticed 
looking at the
Chandra image (Fig.~\ref{xrayimage}a).
To study the radial spectral properties, we subdivided the emission from the galaxy into 25 annuli centered on the
X-ray peak.
We extracted a spectrum from each annulus (excluding the point sources) using
$specextract$, which generates source and background spectra building appropriate RMFs and ARFs. 
The background is taken from the re-normalized blank field observations using the same region as the 
source.
The spectra were analyzed using XSPEC v12 \citep{arnaud96} and fitted with a 
single-temperature {\tt APEC} model \citep{smith01}, combined
with the XSPEC {\tt projct} deprojection model.
% in which the 
%ration
%between the elements is fixed to the solar value as in Anders \& Grevesse (1989). 
%However, as explained in \S1, these
%values for the solar metallicities have more recently been superseeded by the new values by
%Grevesse \& Sauval (1998) and A05. For clarity and completeness, we 
%also performed the fits using solar abundances by A05.
The free parameters in the model are 
the gas temperature $kT$, the gas metallicity $Z$ and the normalization. 
The spectral band considered is 0.5-4 keV. 
%We choose not to consider the data below 0.5 keV because of 
%uncertainties
%in the ACIS calibration below that energy.
All the spectra were rebinned to have at least 20 counts per bin.
% and to be 
%able to use the $\chi^2$ statistics. 
%Although because of our low energy cut at 0.6 keV we become less sensitive in measuring the
%$N_H$, we avoid to introduce systematics due to the uncertainties in the calibration 
%of the instrument at soft energies. 
We have measured $N_H$ from the X-ray data, finding it consistent (within 1$\sigma$) 
with the Galactic value
along the line of sight, as derived from radio data 
\citep[$1.9\times10^{20}$ cm$^{-2}$;][]{stark92}.
The de-projected temperature profiles of NGC~4636 
are shown in Figure~\ref{ktz}a. No abundance profile trend was detected, we therefore decided to fix the
abundance at $Z=0.5$ times solar. This is in agreement with the findings of \citet{osullivan05}, who
also found no trend in $Z(r)$ for the azimuthally-averaged data. Interestingly, they did however find 
a strong
azimuthal abundance variation.  The temperature shows clear evidence for a decline in the
center where $kT\sim0.5$~keV, and increases toward the outskirts of the galaxy ($kT\ga 0.8$~keV). 
Several substructures, due to the
complex X-ray morphology of the galaxy, are observed.
The electron density profile derived from the deprojection (Fig.~\ref{ktz}b) shows instead a strong central
peak and a `plateau' coincident with the one observed in the surface brightness profile located at 
2~kpc$\la r\la$4.5~kpc from the center.

\subsection{The X-ray Bubbles}
The complex X-ray morphology of NGC~4636 is mainly constituted by bubbles and cavities, whose properties cannot
be determined with a simple radial analysis.
Therefore, we studied the features observed in the X-ray image in detail, performing
a spatially resolved spectral analysis on these regions, looking for evidence of the feedback from the central
AGN in the surrounding hot interstellar and intergalactic medium.

\subsubsection{The SW Bubble}

\begin{figure}
%\epsscale{.80}
\plotone{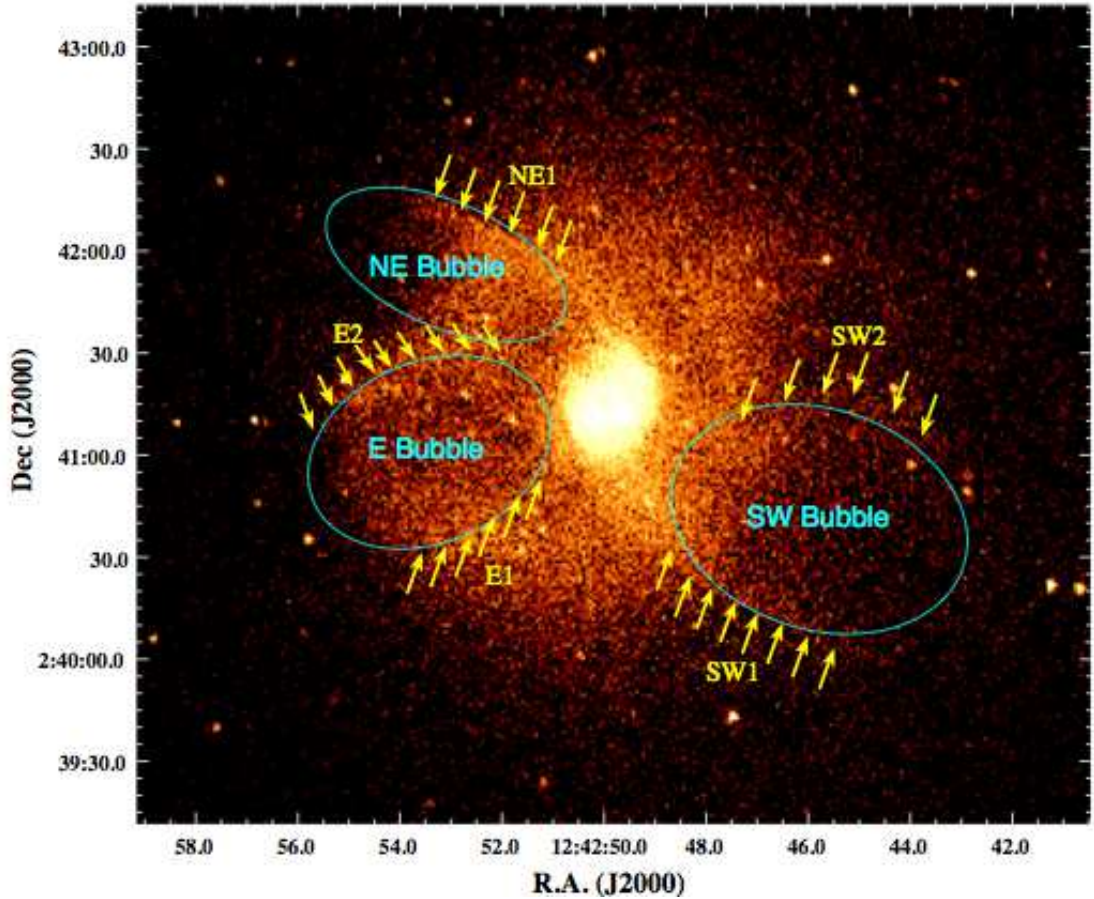}
\caption{Chandra ACIS-I+ACIS-S image of NGC~4636 (0.5-3 keV band). The three bubble-like features
detected are labelled and identified by cyan ellipses. The yellow arrows are pointing toward 
the detected rims of the bubbles.\label{xrayarrows}}
\end{figure}

The most prominent features observed in the Chandra image are clearly the two quasi symmetric 8~kpc
long X-ray arm-like structures, already observed by \citet{jones02} and by \citet{osullivan05}. The SW arm (SW1) 
is however part of a cavity
which extends at least $\sim5$~kpc in radius to the North (Fig.~\ref{xrayarrows}), although the northern
boundary of the bubble (SW2) is not as sharp and well defined as the southern one.
We performed a spectral analysis in a strip perpendicular to the SW arm-like feature, dividing
the strip into rectangles $\sim0.5^\prime$ wide (Fig.~\ref{xrayspecreg}), 
to look for variations in temperature or metal abundance. 
Although the metallicity does not vary significantly
across the bubble rim, a sharp variation in the temperature was clearly detected coincident with SW1 
(Fig.~\ref{SWshock}) 
showing a temperature decrease from $kT\sim0.75$~keV to $kT\sim0.64$~keV, well above
the measurement errors ($\sim0.01$~keV). A temperature jump was not observed across SW2 most likely because
of the complicated geometry of the X-ray emission. Indeed, SW2 seems to be partially embedded in another
bubble-like feature located just North of it.\\

\subsubsection{The NE and the E Bubble}

\begin{figure}
%\epsscale{.80}
\plotone{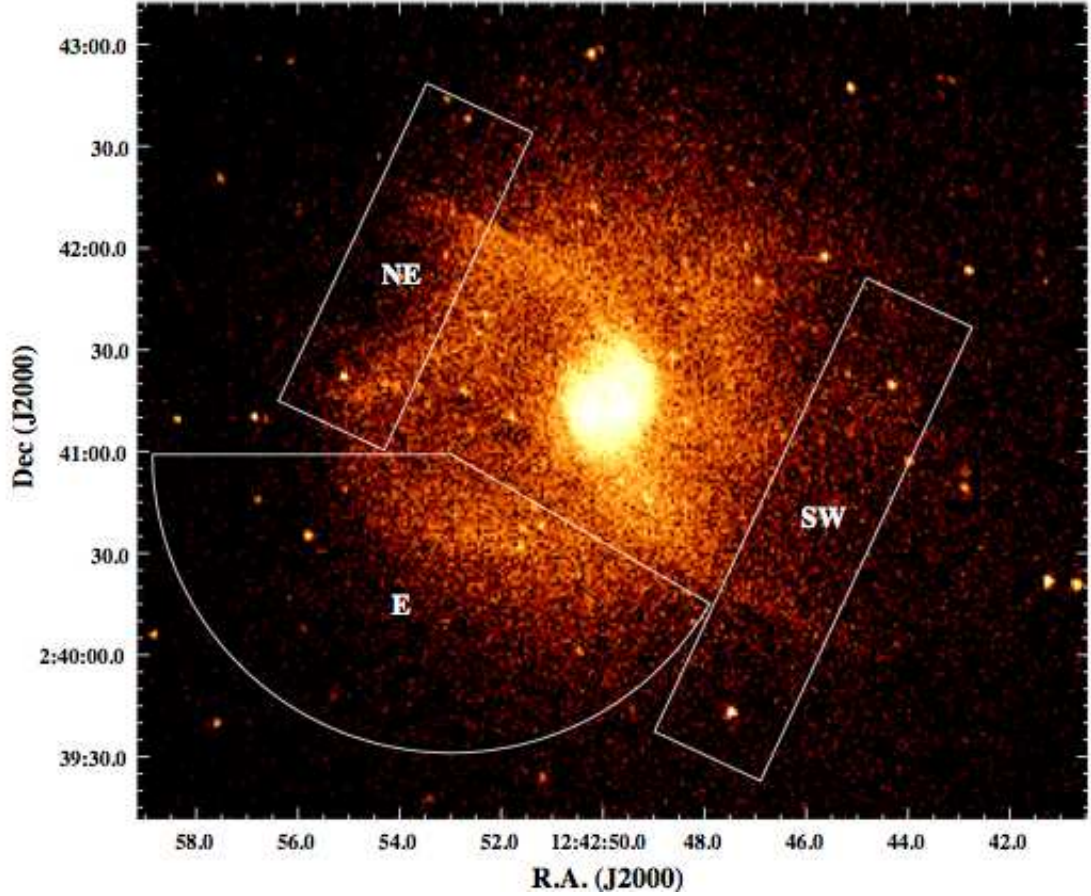}
\caption{Chandra ACIS-I+ACIS-S image of NGC~4636 in the 0.5-3 keV band. The regions used to extract the spectra for
the analysis of the three bubble-like features are shown and labelled in white.\label{xrayspecreg}}
\end{figure}

The symmetry of the X-ray arm-like features is highly suggestive of the presence of a symmetric bubble NE of the nucleus
of NGC~4636. However the southern rim of the bubble is not clearly visible and it looks instead to be embedded
in another round shaped bubble located to the East of the nucleus. If we examine the surface brightness profile of the
NE cavity we find a shape which is very similar to the SW bubble (Fig.~\ref{NEshock}a). Performing a spectral analysis across the
northern cavity rim NE1 (Fig.~\ref{xrayspecreg}), 
we also observe a temperature jump across the rim (Fig.~\ref{NEshock}b). However,
the scenario in this part of the galaxy is more complex because of the presence of an additional feature just East of the
nucleus with the shape of another cavity. This cavity looks less elongated than the other two cavities observed.
Therefore, we preferred to analyze the temperature variation across its southern rim E1 (the northern boundary E2 could be
contaminated by the presence of the NE cavity) using partial annuli instead of rectangular regions (Fig.~\ref{xrayspecreg}).
The surface brightness profile has a shape similar to the one across the NE cavity, while the temperature profile shows
a temperature jump similar to the one observed in the other two cavities (Fig.~\ref{Eshock}).

%\begin{figure}
%%\epsscale{.80}
%\plottwo{kTprof_SE_deproj.ps}{neprof_SE_deproj.ps}
%\caption{{\it Left:} Deprojected temperature radial profile along a wedge located SE of the nucleus of NGC~4636. 
%The direction of the wedge was chosen to avoid any X-ray feature such as cavities or bubbles. {\it Right:} 
%Density radial profile along the SE wedge.\label{SEradial}}
%\end{figure}

\subsubsection{The origin of the cavities: a simple shock model}
The most likely scenario for the origin of the cavities observed in the X-ray morphology of NGC~4636
is that they were the result of successive outbursts of the central AGN.
% similarly to what happened in the 
%case of the elliptical galaxy M~84 \citep{finoguenov08}.
In this scenario the jet propagated rapidly from the center, creating a long thin cavity which then inflated in
all directions. Perpendicular to the axis of the jet, the expansion has an approximate cylindrical symmetry,
motivating our choice to use rectangular regions and therefore assume cylindrical symmetry.
The expanding cavities drive shocks into the surrounding gas.

A 1-dimensional, cylindrically symmetric, time-dependent hydrodynamic
model was used to investigate the properties of these shocks for the
SW bubble.  
In this model the sound speed in the relativistic gas that fills the cavity (the piston that drives the shock) 
is assumed very high, keeping the pressure in the piston nearly uniform. The ratio of the pre-shock
pressure to the post-shock pressure determines the strength of the shock. As a result, the shock is weakest 
(slowest) in the region close to the AGN and fastest in the region farthest from the AGN. 
%The
%pressure required to drive the shock observed at the middle of the cavity now is lower than the ambient 
%pressure close to the AGN, so the piston can no longer drive the shock towards the AGN. Indeed,
%the end of the cavity close to the AGN is being pushed outward, as denser gas falls inward and displaces it.

\begin{figure}
%\epsscale{.80}
\plottwo{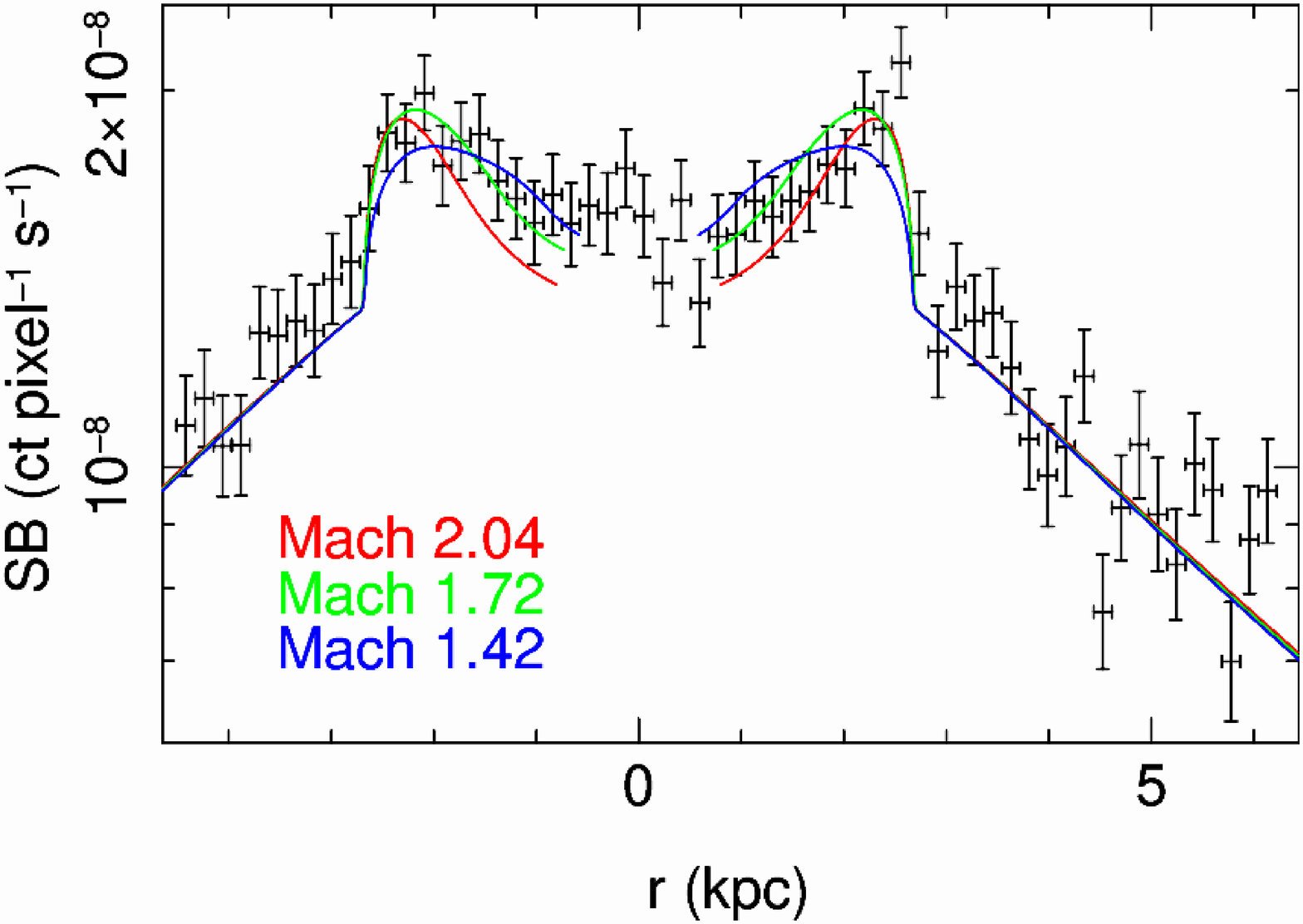}{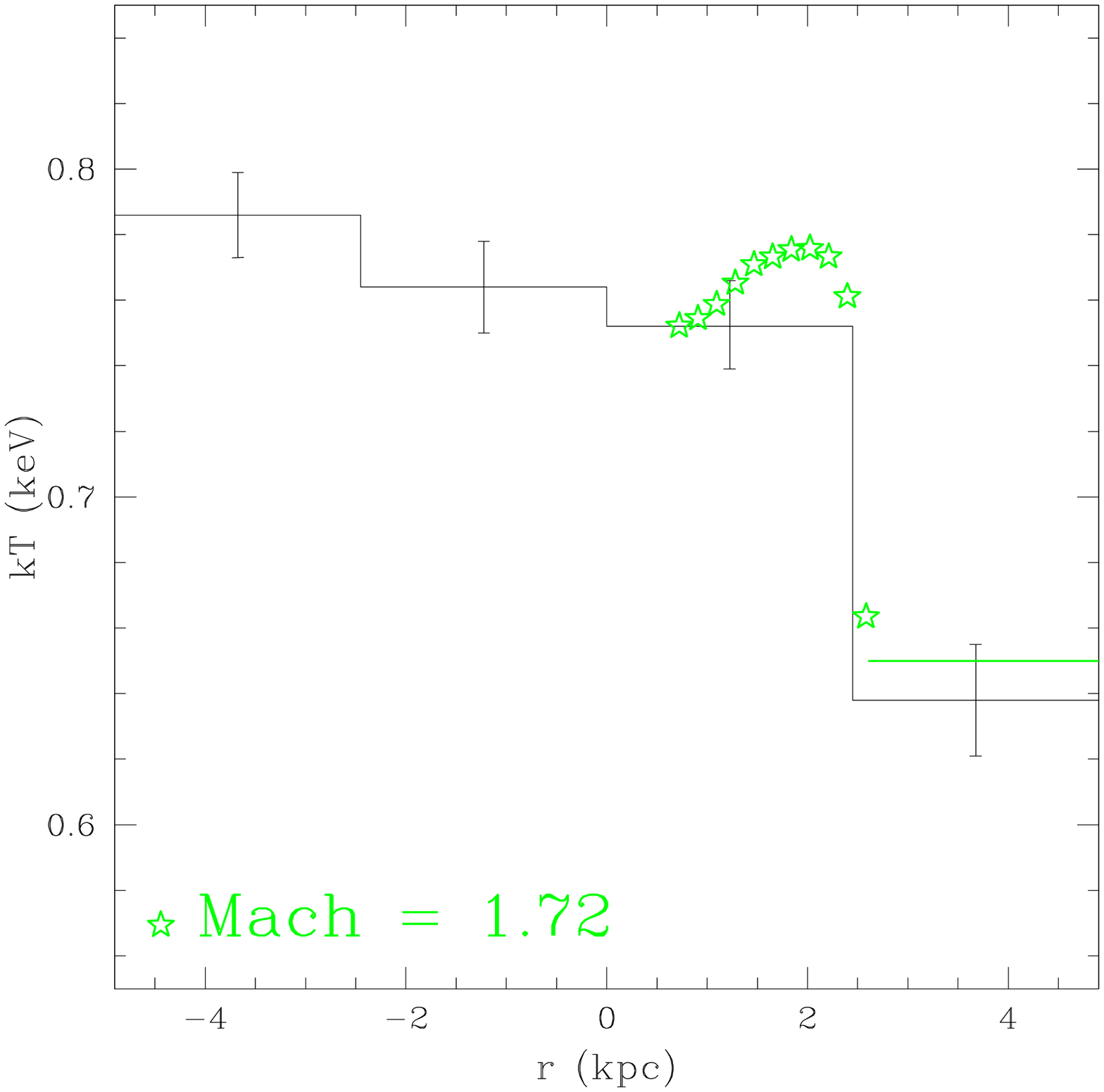}
\caption{{\it Left:} Surface brightness profile perpendicular to the shock front for the SW bubble. The three
colored solid lines represents the prediction from a numerical hydro-dynamical shock model at different
Mach numbers for the shock. The best fit shock model to the observed data has $M=1.72$. {\it Right:} 
Temperature profile across the southern rim of the SW bubble. The temperature jump is consistent with
the predictions of a shock model with $M=1.72$. The zero points in $r$ are coincident in the two panels and the values of $r$ increase 
toward the South.\label{SWshock}}
\end{figure}

\begin{figure}
\plottwo{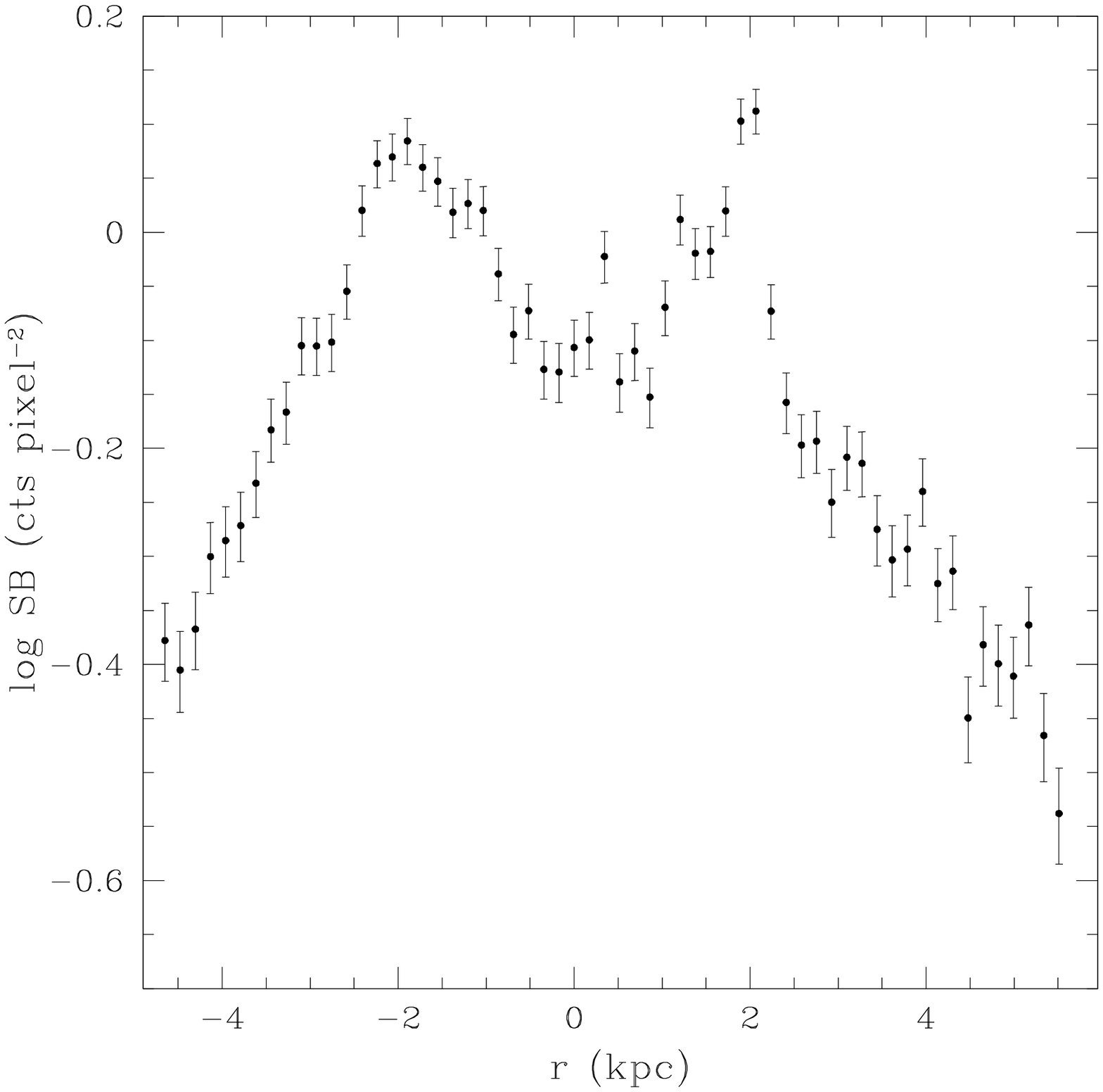}{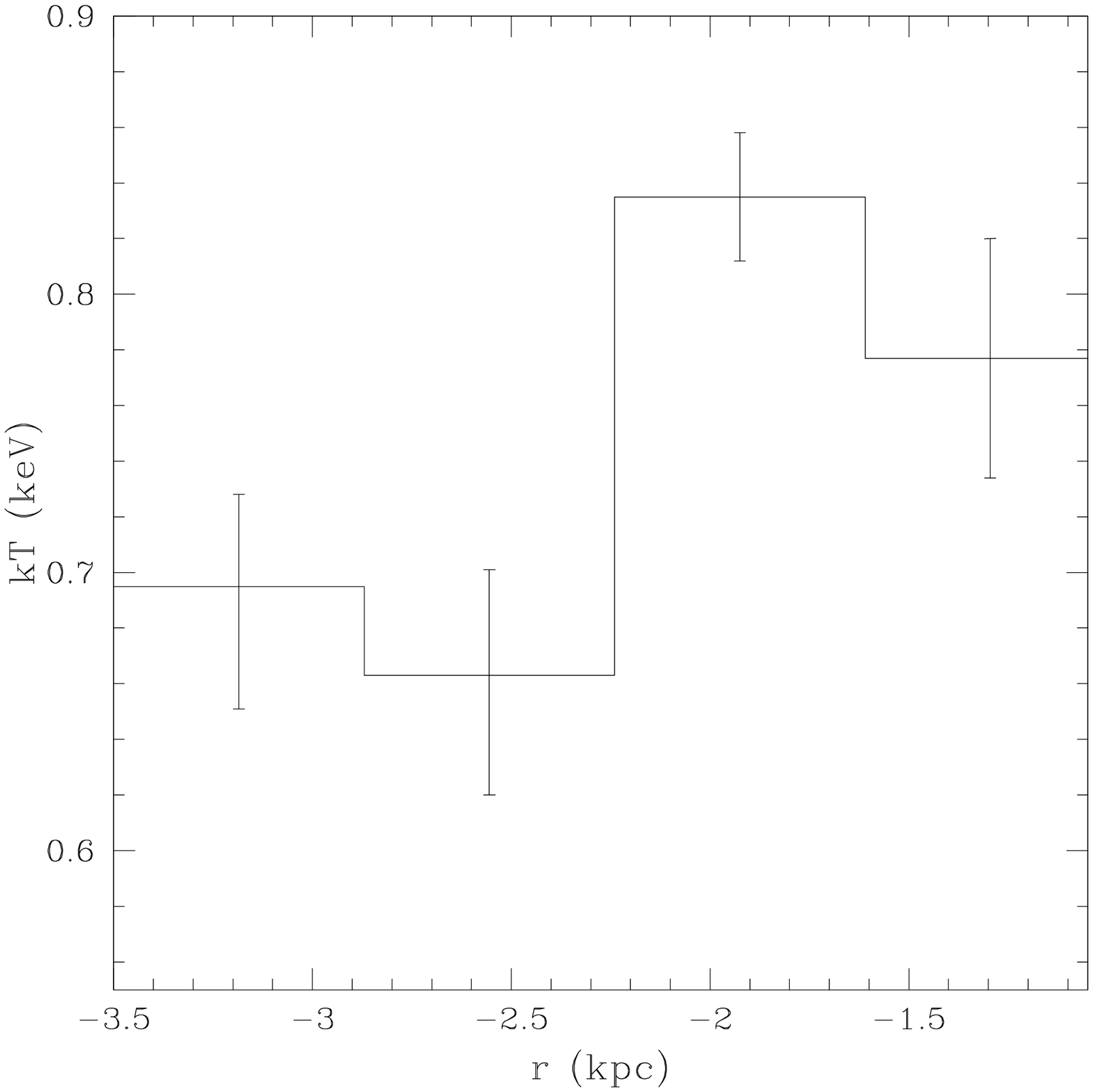}
\caption{{\it Left:} Surface brightness profile perpendicular to the shock front for the NE bubble. {\it Right:} 
Temperature profile across the northern rim of the NE bubble. The shape of the surface brightness profile is very
similar to the one observed in the SW bubble, moreover a temperature jump is observed in coincidence with the
northern cavity rim (NE1). The zero points in $r$ are coincident in the two panels and the values of $r$ increase 
toward the South.
\label{NEshock}}
\end{figure}

\begin{figure}
%\epsscale{.80}
\plottwo{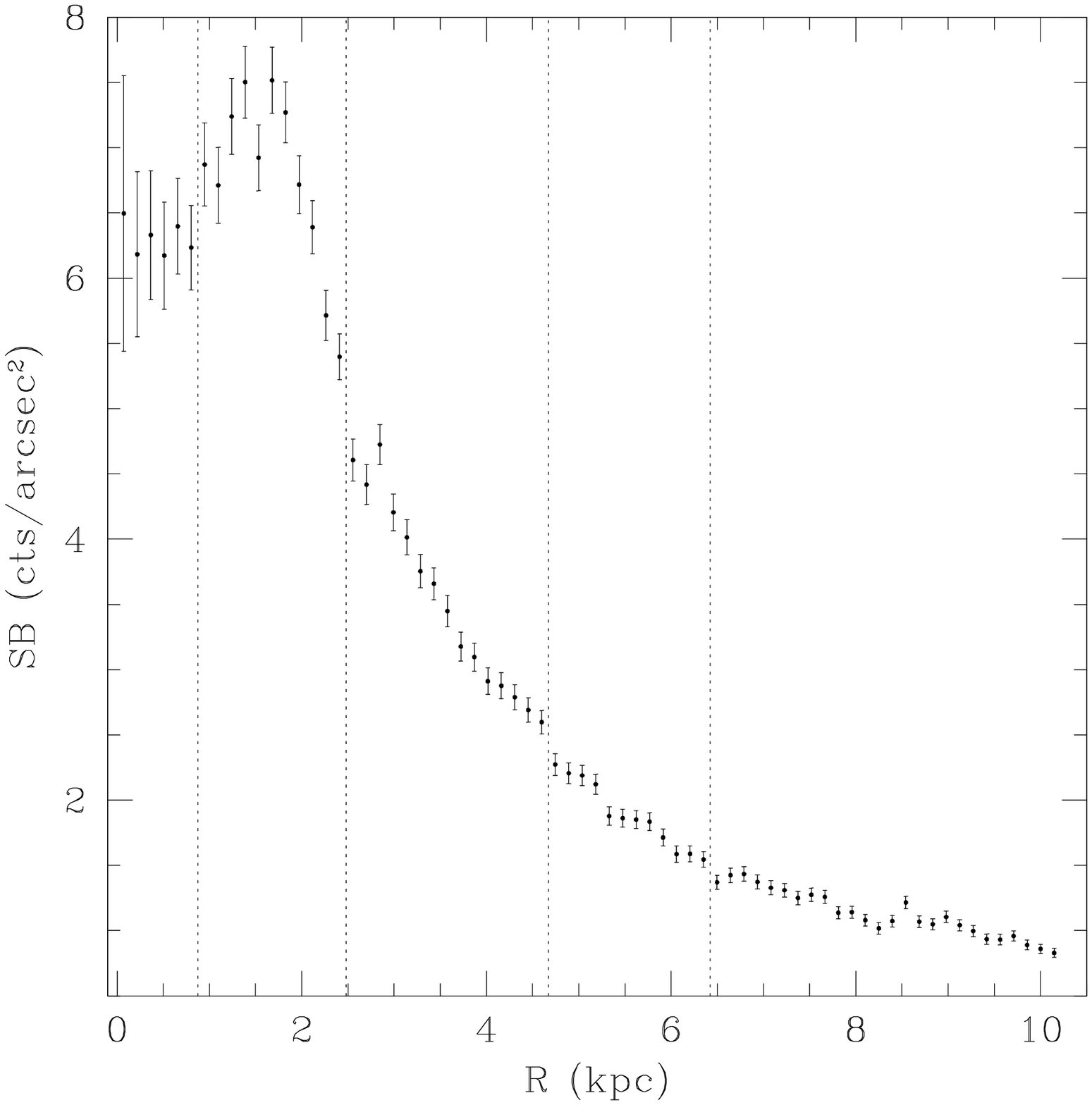}{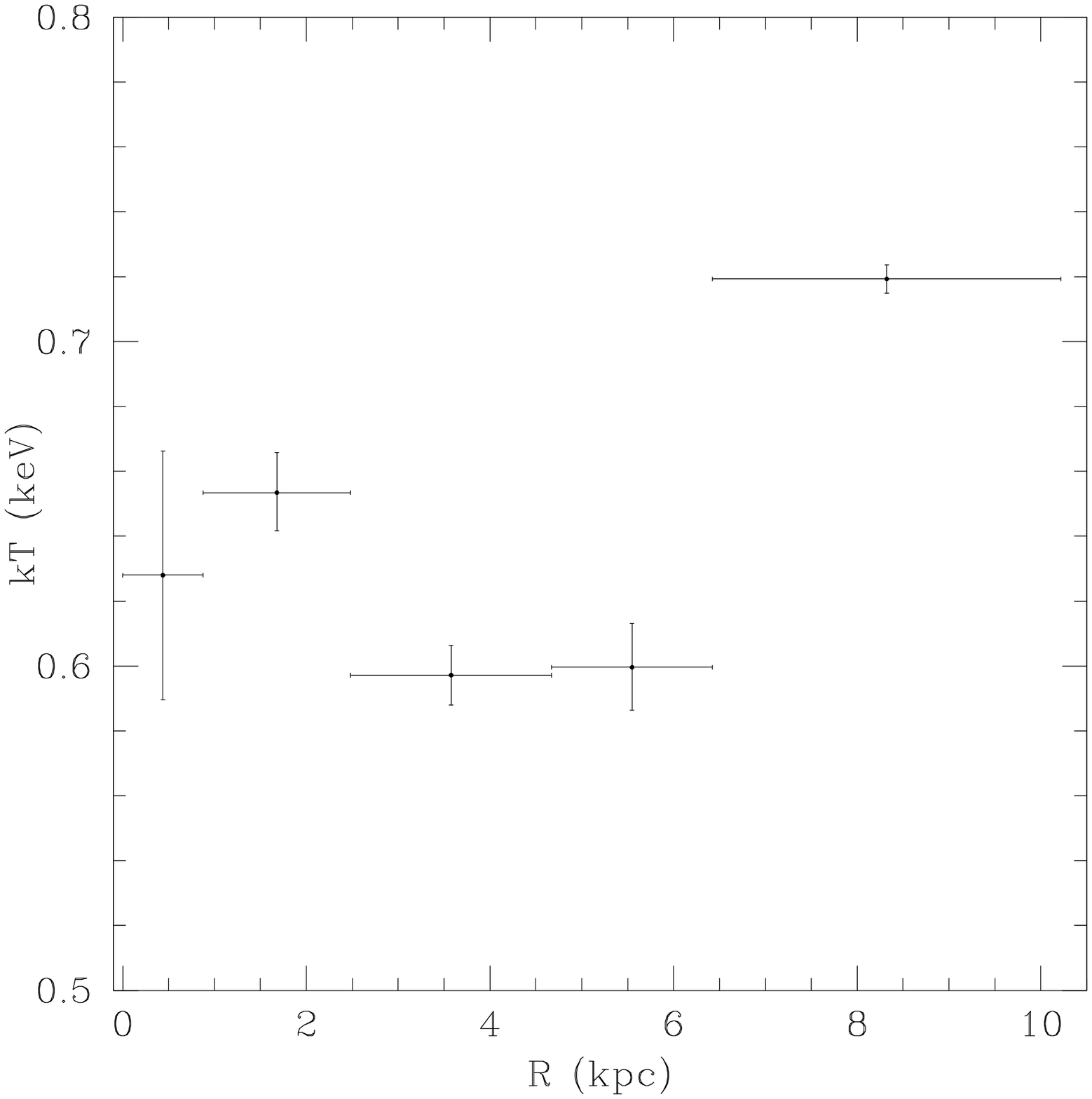}
\caption{{\it Left:} Surface brightness profile across the southern rim for the East bubble. The dashed vertical lines
delimitate the regions used for the spectral analysis. {\it Right:} 
Temperature profile across the southern rim of the E bubble. The shape of the surface brightness profile is 
analogous to the one observed in the other two bubbles and a temperature jump is observed in coincidence with the 
southern rim (E1) at $\sim2.5$~kpc from the center of the bubble.\label{Eshock}}
\end{figure}

The undisturbed gas was assumed to be isothermal, with electron
density $n_{\rm e,gas} \propto r^{-\eta}$, where $r$ is the
distance to the AGN.  Corresponding to this, the surface brightness
profile of the undisturbed gas has the form, ${\rm SB} \propto
\rho^{1 - 2 \eta}$, where $\rho$ is the projected distance from the
AGN.  The power law index $\eta$ was determined by fitting the SB
profile in the region adjacent to the SW arm, outside the shock,
giving $\eta = 1.3$.  Taking a cut perpendicular to the radius at a
distance, $d$, from the AGN, gives the electron density profile
$n_{\rm e,gas} \propto (d^2 + \varpi^2)^{- \eta/2}$, where $\varpi$
is the cylindrical radius measured from the axis of the lobe.
Using the distance, $d$, corresponding to $1.2'$, this was the form
adopt for the density profile of the undisturbed gas in the
cylindrical flow model.  The initial gas distribution was made
hydrostatic by imposing a static gravitational field on the model.
Shocks were initiated by depositing a large amount of energy on the
symmetry axis and their evolution was followed using the
cylindrically symmetric hydrodynamic code until the shock had
expanded to the observed size. Surface brightness profiles for
the model shocks were determined by projecting the axially symmetric
models onto the sky, using the density and temperature dependent 0.3
-- 1.8 keV ACIS-I response.  The temperature scale was
determined by assuming a preshock temperature of 0.65 keV.  The left
panel of Fig.~\ref{SWshock} shows the surface brightness profile of
the SW shock, perpendicular to the radius, in the region shown in
Fig.~\ref{xrayspecreg}, together with model surface brightness
profiles for Mach numbers of 1.42, 1.72 and 2.04 (computed
assuming $\gamma = 5/3$ in the gas). Based on its
surface brightness profile in the vicinity of the shocks, the Mach
1.72 shock model provides the best fit.

The temperature jump expected for a Mach 1.72 shock is a factor of
1.74.  For a preshock temperature of 0.65 keV, that would make the
postshock temperature $\simeq 1.13$ keV.  However, as observed, gas
behind the cylindrical shock is projected onto the surrounding
unshocked gas.  The shocked gas also expands adiabatically behind the
shock, reducing its average temperature.  As a result, the observed
temperature profile shows a considerably lower temperature jump than
expected for a Mach 1.72 shock.  Complicating matters further, fitted
temperatures for multiphase gas are not readily determined from the
emission measure distribution \citep{vikhlinin06}.  In view of this,
projected emission measure distributions were determined for the Mach
1.72 model, assuming a preshock temperature of 0.65 keV.  These were
used to construct multiphase spectral models in XSPEC and the
resulting models were fitted to a single temperature model 
to determine the observed temperature
profile corresponding to the Mach 1.72 shock model.  The results are
plotted as stars together with the observed temperature profile in the
right panel of Fig.~\ref{SWshock}.  Within the errors, both the
temperature and density profiles of this region are consistent with
the Mach $\simeq1.7$ shock model.

The calculations of the physical parameters derived from the model were performed for the SW bubble.
The three cavities present similar temperature jumps and surface brightness profiles, so similar
physical parameters are expected.
%The shock strength was determined by matching the observed surface brightness perpendiculary to the cavity rims with the projected
%model (Fig.~\ref{SWshock}a). A Mach number $M\sim1.7$ is found to be the the best match for the observed surface brightness.
%The prediction of a shock model having the aforementioned Mach number is consistent also with the temperature jump 
%observed across the southern boundary (Fig.~\ref{SWshock}b).
From the hydrodynamic model, the age of the shock, $t\sim2\times10^{6}$~yrs, is the time it takes
to expand to its observed size. The age is only sensitive
to the shock strength and the temperature of the unshocked gas. This
age is slightly shorter than the ratio of the shock radius to the
present shock speed, because in the model the shock strength decreases
with time (i.e. the shock expanded faster when it was younger).
The total energy which produced the shock was $10^{56}$~ergs, roughly equal to the enthalpy, 
$H=4pV=7\times10^{55}$~ergs, calculated in the assumption that the bubble is predominantly relativistic 
($\gamma=4/3$). The average mechanical power required to produce the bubble equals to 
$P_{mech}\sim1.6\times10^{42}$~erg s$^{-1}$.

The age of the cavity is shorter than what is usually observed in the so-called 'ghost' cavities
where no evident 1.4~GHz radio emission is observed to fill them. However, the cavities in NGC~4636 are not
really 'ghost' cavities and a weak radio luminosity is indeed expected from the mechanical power derived above.
\citet{bir08} found indeed a correlation (although with a large scatter) between the cavity (jet) power $P_{jet}$ and 
the radio synchrotron power $L_{radio}$, of the form $P_{jet}\sim L_{radio}^{\beta}$.
%, where $0.35\leq \beta\leq 0.70$. 
The cavities observed in NGC~4636 fall in the lower-left corner of the diagram in Figure 5 of \citet{bir08}, with
a ratio $P_{jet}/ L_{radio}\sim10^4$, similar to a few objects in the plot presenting a radio luminosity
four order of magnitudes brighter than NGC~4636. 
There is a large scatter in the correlation, that is not well understood.
Given the associated jet powers, the low level of radio emission from X-ray cavities in some of these
objects is indeed surprising. As discussed in \citet{bir08}, the steep radio spectra in these
systems cannot generally be explained by synchrotron aging or by adiabatic losses. 
Presumably some other loss mechanism is at work. This raises interesting questions on the nature of the
plasma filling these cavities. 
%In the case of NGC~4636, the low radio luminosity observed is consistent with expectations based on this sample.
As stated e.g. in \citet{jetha08}, the nature of this material could be very hot, low-density plasma or
could consist, at least in part, of a relativistic plasma containing relativistic electrons and magnetic
field, similarly to that observed in the lobes of radio galaxies. 
In the case of NGC~4636, however, the complex X-ray morphology and the lack of low frequency radio maps 
(that will be presented in a forthcoming paper; Giacintucci et al. in prep.) does not allow at present to 
constrain the nature of the plasma filling the cavities \citep[as it was done in e.g.][]{bir08,jetha08}.
%If we compare the mechanical power obtained for the cavities with the radio luminosity of the jets creating the cavities
%($L_{1.4\,GHz}\sim1.4\times10^{38}$ erg s$^{-1}$), we observe that the weakness of the radio lobes is just due to a
%scaling factor respect to larger objects such as clusters and groups. In Figure 5 of \citet{bir08} the cavities of NGC~4636
%occupies the lower-left end of the diagram, ho
%\begin{figure}
%\epsscale{.80}
%\plotone{xrayhard.ps}
%\caption{Hard band (2-7 keV) Chandra ACIS-I+ACIS-S image of the core region of NGC~4636. The position of the central 
%point-like radio source is indicated as a white circle. 
%Several point sources are detected and one of the point sources is coincident with the radio position.
%\label{xrayhard}}
%\end{figure}

\subsection{The nuclear point source}

The X-ray morphology of NGC~4636 is characterized by the presence of a dense core having a radius 
of $\sim1$~kpc.
A hard band image of the core (2-7 keV band) reveals the presence of 
several point sources. An X-ray point source is
coincident with the radio nucleus (within 0.5$^{\prime\prime}$) and may be X-ray emission from the
supermassive black hole (SMBH) at the center of the galaxy. We extracted a spectrum of this point source
fitting a power-law model to it in XSPEC. The spectrum has $\sim230$ counts (ACIS-S plus ACIS-I) in total
and the best fit model gives a power-law index $\Gamma=2.5_{-0.5}^{+0.4}$ and
a flux of 2.3$\pm0.4\times10^{-15}$ erg cm$^{-2}$ s$^{-1}$ and of 6.0$\pm1.0\times10^{-15}$ erg cm$^{-2}$ s$^{-1}$ in the
2-10 keV and in the 0.5-10 keV bands, respectively. This corresponds to a total luminosity of 
1.6$\pm0.3\times10^{38}$ erg s$^{-1}$ in the 0.5-10 keV band. This could be the current luminosity of the AGN at the
center of NGC~4636. However, given the low luminosity, the X-ray emission
could be also due to a low mass X-ray binary coincident with the radio nucleus. In the latter case, 
this value can be considered 
as a more stringent upper limit to the X-ray luminosity of the AGN at the center of the galaxy than
the one set by \citet{loewenstein01}.

To derive an estimate of the mass supply rate for accretion on the SMBH for NGC~4636, we can
apply the \citet{bondi52} theory of steady, spherical and adiabatic accretion. This requires $T$
and $n$ at ``infinity'', in practice near the accretion radius (that
in the case of NGC~4636 is $r_A=GM_{BH}/c_s^2\sim 8$ pc, where $c_s$ is
the speed of sound).  Unfortunately the spatial resolution does not allow us to measure the 
temperature and the density at such a small radius. However, we can estimate
the accretion rate, using the temperature and the density profiles measured in
in our radial analysis (\S~\ref{global}).
The accretion rate is given by the formula
$$
\dot{M}_{Bondi} = 8.4 \times 10^{21} \;
M_8^2\;\;T_{0.5}^{-3/2} \;n_1 \; \hbox{$\thinspace 
g\; s^{-1}$}, 
$$
where $M_8$ is the SMBH mass in units of $10^8\; M_{\odot}$ \citep[for
NGC~4636 $M_{BH}\simeq7.9\times10^7\; M_{\odot}$,][]{merritt01}, $T_{0.5}$ is the
temperature in units of 0.5 keV and $n_1$ is the density in units
of 1 cm$^{-3}$.
% [see, e.g., eq.  (3) of \citet{dim03} ].
Assuming that the density $n\propto r^{-1}$, from the thermal component normalization
of the spectrum in the inner
5$^{\prime\prime}$ ($\sim0.35$~kpc), we obtain that the density at the accretion radius is
$n\sim5.5$~cm$^{-3}$. The temperature profile was instead fitted with a second order
polynomial, giving a value at the accretion radius of $kT\sim0.49$~keV.
Using this values of $n$ and $kT$, we find that
$\dot{M}_{Bondi}\simeq 4.7\times10^{-4} M_{\odot}$ yr$^{-1}$. 
If at very small radius this gas joins an accretion
disc with a standard radiative efficiency $\eta \sim 0.1$, as in
brighter AGNs, it should produce a luminosity:
$$
L_{acc}=\eta \dot{M}_{Bondi}
c^2 \simeq 2.7 \times 10^{42} \hbox { erg s$^{-1}$},
$$
at least four orders of magnitude higher than the
observed X-ray luminosity, pointing toward a highly inefficient accretion scenario. \\

\subsection{The bright core of NGC~4636}

\begin{figure}
%\epsscale{.80}
\plotone{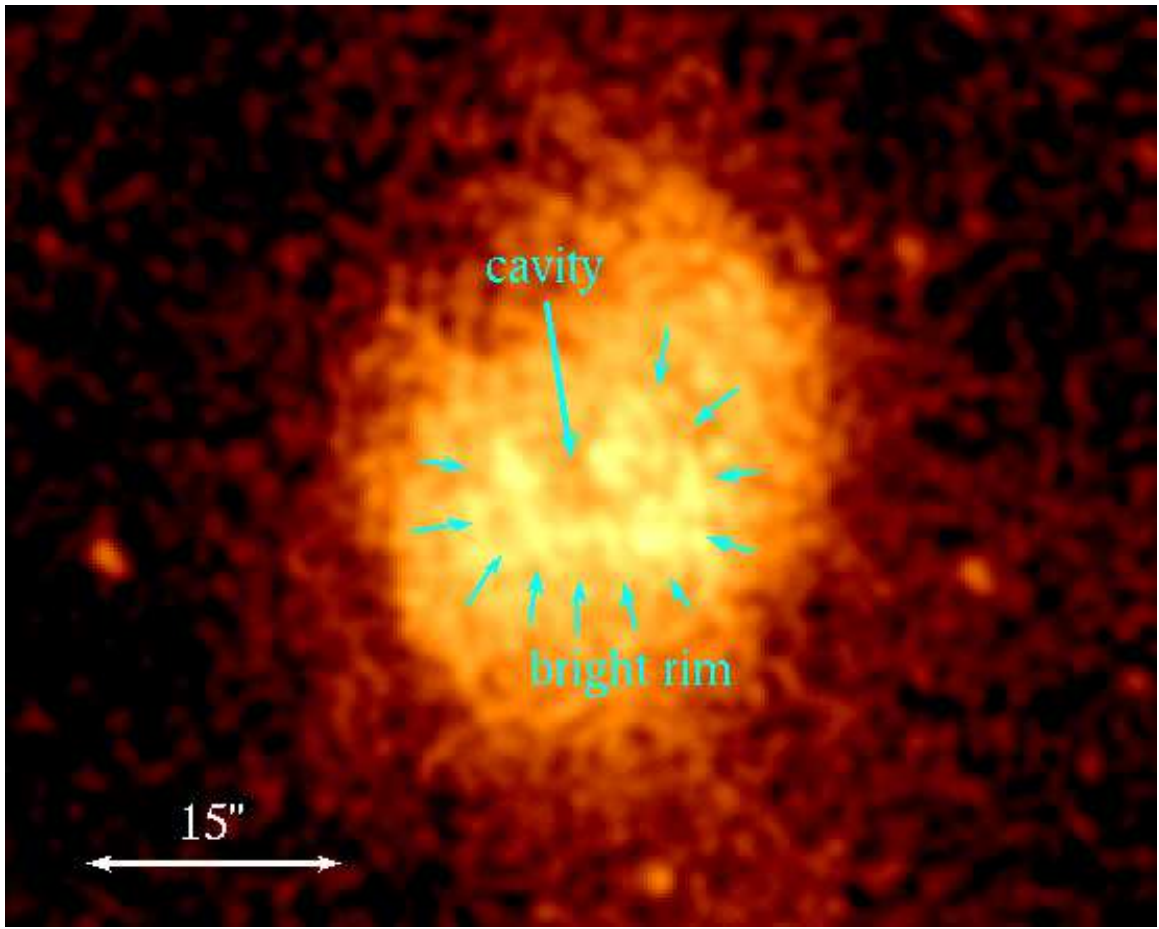}
\caption{Chandra ACIS-I+ACIS-S image of the core of NGC~4636 (0.5-3 keV band). 
A cavity in the galaxy core is well visible and it is surrounded by a U-shaped brighter rim.\label{xraycore}}
\end{figure}

\begin{figure}
%\epsscale{.80}
\plotone{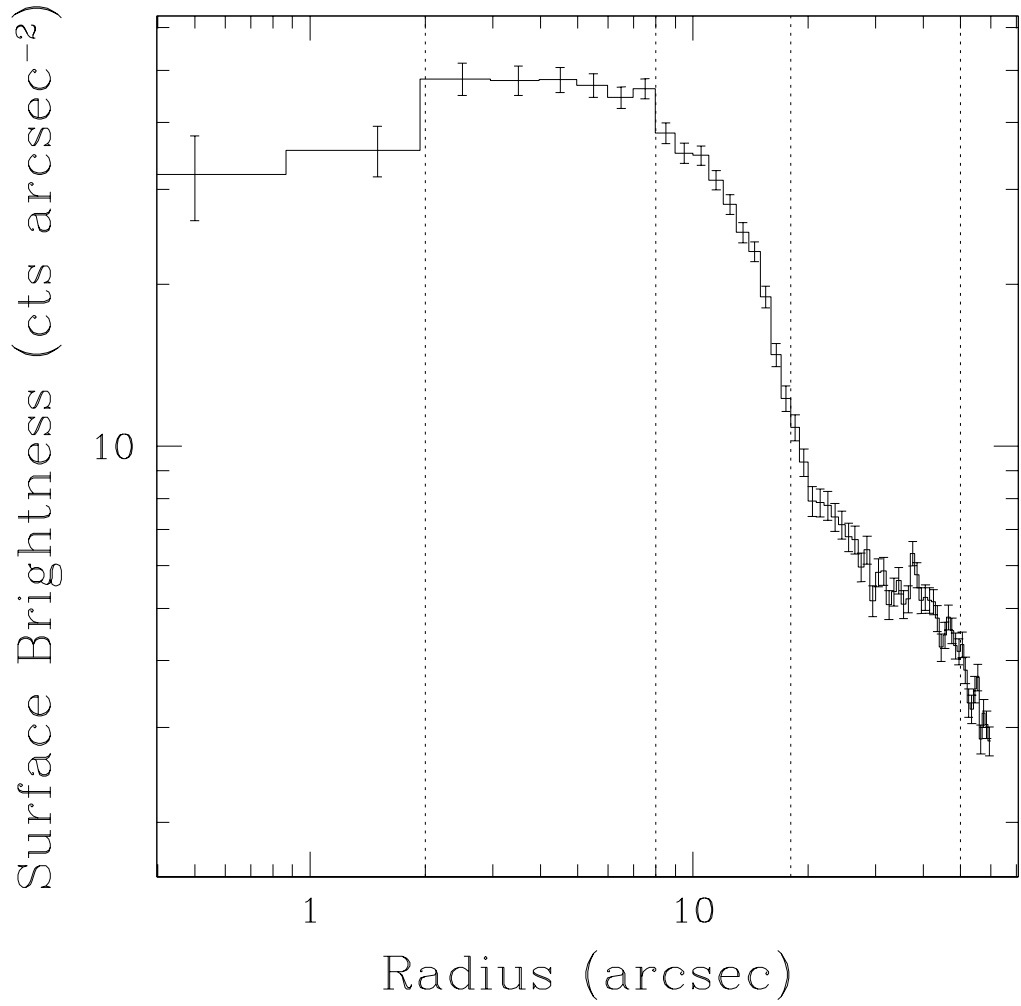}
\caption{Surface brightness profile in a sector west of the nucleus ($315^\circ$--$405^\circ$), extracted in the
0.5-3 keV band. 
A depression in the center is well visible, followed by a flat region and then a sharp decline of the surface
brightness. The dashed vertical lines
delimitate the regions used for the spectral analysis.\label{sb315405}}
\end{figure}

\begin{figure}
%\epsscale{.80}
\plottwo{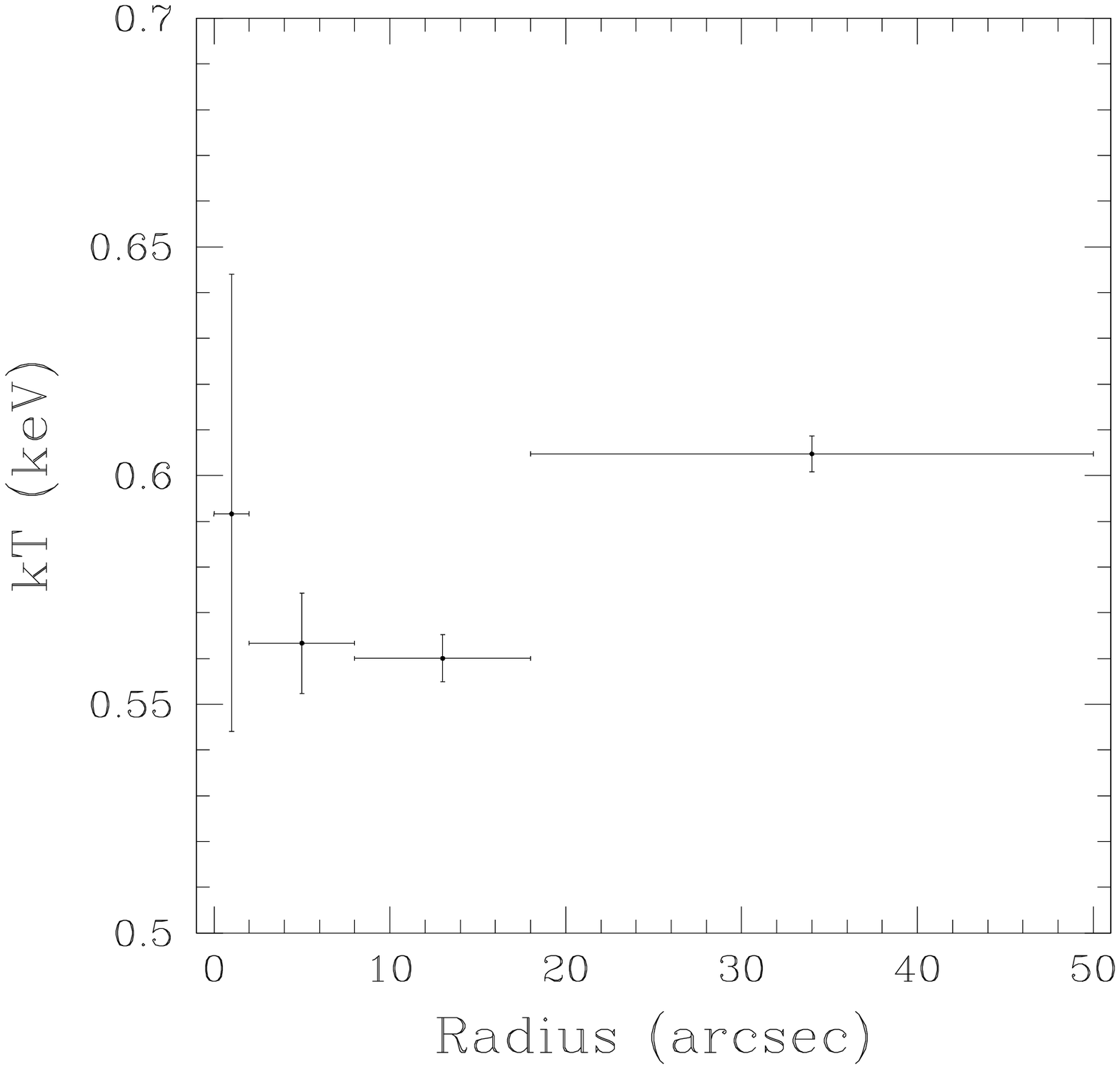}{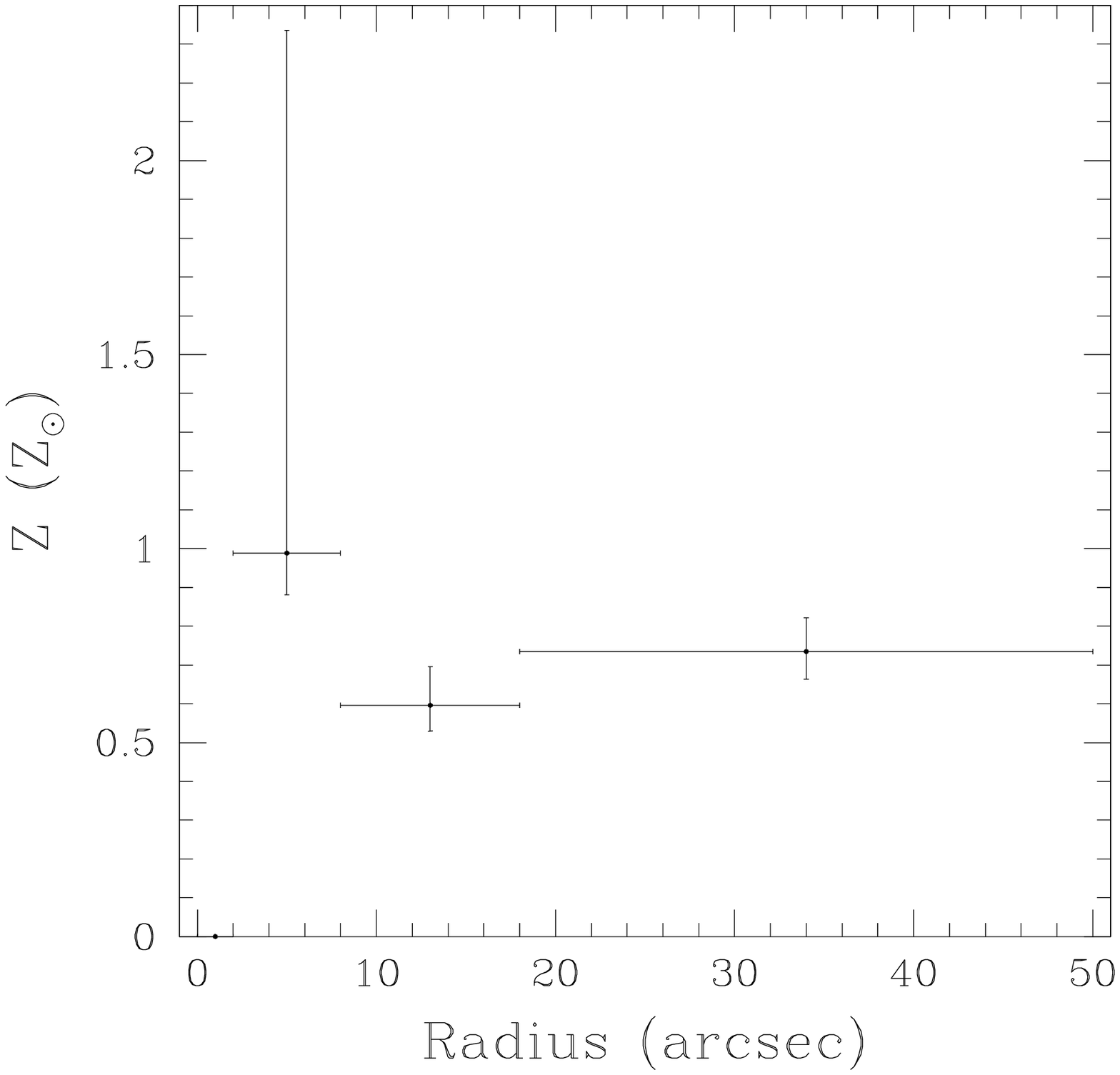}
\caption{{\it Left:} Temperature profile in the $315^\circ$--$405^\circ$ sector. {\it Right:} Abundance
profile in the $315^\circ$--$405^\circ$ sector. Both the temperature and the abundance profile are
quite flat and do not present evidence of strong gradients with the radius.\label{kTZcore}}
\end{figure}

The average density in the core is $\sim0.1$ cm$^{-3}$ and the thermal pressure (computed as 2.2$nkT$, in the
case that $n=n_p$) is 
$\sim9.7\times10^{-11}$
dyn cm$^{-2}$. Several substructures are visible as well as can be seen from Figure~\ref{xraycore} where
an X-ray image of the central part of the galaxy is shown. 
This X-ray image (0.5-2 keV band) shows a cavity in the center of the core,
%which seems to be `filled' by the radio jets, which extends toward the beginning of the two main cavity-like
%structures (NE and SW bubble).
although it is not clear, however, if we are dealing with a cavity or with 
a U-shaped enhancement in the surface brightness of the core. Interestingly, 
the position of this cavity is coincident with the inner radio jet detected at 1.4~GHz.\\
We performed a finer spatial and spectral analysis of the core region, considering the X-ray emission west 
of the galaxy center ($315^\circ$--$405^\circ$), where a sharper decline in surface brightness is detected.
This sector was chosen also because the discontinuity matches the curvature of a wedge centered on the nucleus.
The surface brightness profile in this sector is shown in Figure~\ref{sb315405}.
In this profile, four regions with different behaviours can be identified easily. At the center a depression in the surface
brightness (visible in Figure~\ref{xraycore}) is apparent. Outside the center 
($2^{\prime\prime}<r<8^{\prime\prime}$) we observe a flat region, followed by a sharp decline at 
$8^{\prime\prime}<r<18^{\prime\prime}$. At $r>18^{\prime\prime}$ the surface brightness flattens again then
declines more gradually.
We performed a spectral analysis in these four distinct regions to investigate variations in the temperature or
in the abundance of the X-ray emitting gas. Because of the complexity of the emitting region and the number of counts
detected we could not perform a meaningful deprojection, and we analyzed only the properties of the hot gas
projected along the line of sight. A simple thermal model, plus a power-law to take into account the contribution of
unresolved point sources, was considered (XSPEC model: {\it wabs(apec+powerlaw)}). The abundance was left free to vary
in all the regions but the central ($r<2^{\prime\prime}$), where it was fixed to 0.5 times solar because of the
lower number of counts. Although a one temperature fit does not always account for the complexity of the spectra 
($\chi^2_\nu\geq1.3$), it gives an indication of the general temperature and metal abundance trend.
Temperature and abundance profiles for the sector are plotted in Figure~\ref{kTZcore}.
The temperature is constant at $2^{\prime\prime}<r<18^{\prime\prime}$ ($kT=0.56\pm0.01$~keV) and it 
is slightly increasing right
outside the core ($kT=0.60\pm0.01$~keV). The sharp decline in surface brightness (a factor of at least 5 over a range of 
10$^{\prime\prime}$) is indicative of a drop in the density of a factor of at least 2. The drop in density is 
clearly not accompanied by a corresponding variation in the temperature, thus if the thermal pressure is
the only acting force in the central region of NGC~4636, the bright core is not in pressure equilibrium with 
the ambient gas and cannot be a long lived structure. 
At least two possible scenarios can be invoked to explain the presence of the bright core. 
One scenario involves the presence of the radio lobes which created the cavities observed in NGC~4636, 
where the pressure induced by the relativistic plasma 
could restore the pressure equilibrium between the core and the ambient gas.
In the other scenario the enhancement in brightness of the core region could be explained by an abundance gradient
\citep[as in e.g. NGC~507;][]{kraft04}. In Figure~\ref{kTZcore}, a gradient in abundance between the flat region just outside
the nucleus ( $2^{\prime\prime}<r<8^{\prime\prime}$), where $Z=1.0_{-0.1}^{+1.3}\:Z_\odot$, and the adjacent regions, where
$Z\sim$0.6--0.7 $Z_\odot$, is observed. This difference would not be enough to balance the pressure 
since the density is proportional to $Z^{-1/2}$, however the error bars are quite large and the fits demand for a more complex
spectral model ($1.3\leq\chi^2_\nu\leq1.7$).
We performed a simple calculation to measure the diffusion timescale for Iron (and consequently for the
other elements) to check whether the abundance gradient necessary to balance the pressure in the core could be long lived. 
For a thickness of the transition region (between the peak in surface brightness and the flattening) 
of $\sim1$~kpc, assuming Spitzer type collision cross-sections for Fe, a density of 0.1~cm$^{-3}$, 
and a gas temperature of $kT\sim0.5$~keV, the timescale for Fe
diffusion is a few $10^{13}$~yrs (i.e. several orders of magnitudes longer than a Hubble time).
For the lower atomic weight elements the diffusion timescale would be shorter, going approximately
as the atomic weight cubed (e.g. a factor of $\sim40$ times shorter for O than
Fe). However, also considering lower atomic weight elements and all the possible uncertainties
in the measure of the abundance (mostly due to the strong correlation between $kT$, $Z$ and flux in
cool systems as NGC~4636), of the density and of the thickness of the transition region,
the bright core could be very long lived if produced by
an abundance gradient (and diffusion is covered by Coulomb collisions).

%Avoiding the complications introduced by the presence of the cavities
%we performed a radial spectral analysis of the X-ray emission SE to the galaxy center, where no cavity is present.
%A deprojection technique similar to the one applied in \S~\ref{global} was used to subtract the contribution of the
%outer shells to the inner shells.
%We did not find any radial variation in the metal abundance, therefore we decided to fix it at $Z=0.5$ times the solar value.
%The temperature and density profile of the SE wedge are shown in Figure~\ref{SEradial}.
%In these plots the inner two bins are referred to the 1~kpc-sized core observed in the X-ray image.

\section{Conclusions}

In this paper, we presented Chandra ACIS-I ($\sim150$~ks integration time) and ACIS-S
($\sim40$~ks) observations of the X-ray luminous elliptical 
galaxy NGC 4636, located in the outskirts of the Virgo cluster.
The main results can be summarized as follows:\\
\begin{itemize}
\item A soft band (0.5-2 keV) image shows the presence of a bright core in the center surrounded by an extended 
X-ray halo and two prominent quasi-symmetric, 8 kpc long, arm-like features, both of which are rims of 
ellipsoidal bubble-like structures (NE and SW bubbles). An additional bubble-like feature (E bubble) 
is detected just south of the NE bubble.
\item We found that the temperature profiles across the bubble edges are sharp
and characterized by the presence of temperature jumps of about 20-25\%. On the other hand, the
metallicity across the rims is constant.
\item Analyzing the core region of the galaxy we discovered the presence of a cavity around a weak two-sided
radio jet ($L\sim1.4\times10^{38}$ erg s$^{-1}$). Several point sources are detected in the hard band (2-7 keV)
in the core region. One of the point sources is coincident with the central radio source; its low X-ray luminosity
of 1.6$\pm0.3\times10^{38}$ erg s$^{-1}$ may be consistent with that of a low mass X-ray binary coincident with a
radio nucleus, however this luminosity may at least be considered as an upper limit to the present X-ray
luminosity of the central AGN.
\item We compared the observed temperature and surface brightness profiles across the bubble rims
with numerical hydrodynamic shock models, demonstrating that a scenario where the bubbles were produced by 
shocks can explain the X-ray morphology. 
The shocks were probably driven by energy deposited off-center by jets originating from the central AGN.
A further confirmation of this scenario is the presence of radio jets extending towards the bubbles and 
of a central weak X-ray and radio nucleus, most likely the currently observable signature of AGN activity,
which was more intense in the past.
\item For the SW bubble the shock has a Mach number of $\sim1.7$, an age of $\sim2\times10^6$ yrs, and 
a total energy required to produce the shock of $\sim10^{56}$ ergs.
\item A very bright dense ($n\sim0.1$~cm$^{-3}$) core of $\sim$1~kpc (15$^{\prime\prime}$) radius is observed 
in the galaxy center.
The sharp decline in surface brightness (and consequently in density) in the core is not accompanied by a 
variation in temperature, not
balancing the thermal pressure in the core with the ambient gas. This feature could be however long lived if the relativistic
plasma injected beyond the core by the radio jets
is balancing the thermal pressure or if the surface brightness enhancement is produced by an abundance gradient.
\end{itemize}
%We leave the spectra ungrouped using the Cash statistics to fit the data. 


%% Included in this acknowledgments section are examples of the
%% AASTeX hypertext markup commands. Use \url without the optional [HREF]
%% argument when you want to print the url directly in the text. Otherwise,
%% use either \url or \anchor, with the HREF as the first argument and the
%% text to be printed in the second.

\acknowledgments

We thank J. Vrtilek and E. O'Sullivan for useful comments and discussion.
We thank the anonymous referee for the useful comments and suggestions helpful
to improve the presentation of the results in this paper.

\bibliographystyle{apj}
\bibliography{alesbib.bib}

\begin{thebibliography}{33}
\expandafter\ifx\csname natexlab\endcsname\relax\def\natexlab#1{#1}\fi

\bibitem[{{Anders} \& {Grevesse}(1989)}]{anders89}
{Anders}, E., \& {Grevesse}, N. 1989, \gca, 53, 197

\bibitem[{{Arnaud}(1996)}]{arnaud96}
{Arnaud}, K.~A. 1996, in Astronomical Society of the Pacific Conference Series,
  Vol. 101, Astronomical Data Analysis Software and Systems V, ed. G.~H.
  {Jacoby} \& J.~{Barnes}, 17--+

\bibitem[{{Awaki} {et~al.}(1994){Awaki}, {Mushotzky}, {Tsuru}, {Fabian},
  {Fukazawa}, {Loewenstein}, {Makishima}, {Matsumoto}, {Matsushita}, {Mihara},
  {Ohashi}, {Ricker}, {Serlemitsos}, {Tsusaka}, \& {Yamazaki}}]{awaki94}
{Awaki}, H., {Mushotzky}, R., {Tsuru}, T., {Fabian}, A.~C., {Fukazawa}, Y.,
  {Loewenstein}, M., {Makishima}, K., {Matsumoto}, H., {Matsushita}, K.,
  {Mihara}, T., {Ohashi}, T., {Ricker}, G.~R., {Serlemitsos}, P.~J., {Tsusaka},
  Y., \& {Yamazaki}, T. 1994, \pasj, 46, L65

\bibitem[{{B{\^i}rzan} {et~al.}(2008){B{\^i}rzan}, {McNamara}, {Nulsen},
  {Carilli}, \& {Wise}}]{bir08}
{B{\^i}rzan}, L., {McNamara}, B.~R., {Nulsen}, P.~E.~J., {Carilli}, C.~L., \&
  {Wise}, M.~W. 2008, \apj, 686, 859

\bibitem[{{B{\^i}rzan} {et~al.}(2004){B{\^i}rzan}, {Rafferty}, {McNamara},
  {Wise}, \& {Nulsen}}]{birzan04}
{B{\^i}rzan}, L., {Rafferty}, D.~A., {McNamara}, B.~R., {Wise}, M.~W., \&
  {Nulsen}, P.~E.~J. 2004, \apj, 607, 800

\bibitem[{{Bondi}(1952)}]{bondi52}
{Bondi}, H. 1952, \mnras, 112, 195

\bibitem[{{Br{\"u}ggen} \& {Kaiser}(2002)}]{bruggen02}
{Br{\"u}ggen}, M., \& {Kaiser}, C.~R. 2002, \nat, 418, 301

\bibitem[{{Buote}(2000)}]{buote00}
{Buote}, D.~A. 2000, \apj, 539, 172

\bibitem[{{Cavaliere} {et~al.}(2002){Cavaliere}, {Lapi}, \&
  {Menci}}]{cavaliere02}
{Cavaliere}, A., {Lapi}, A., \& {Menci}, N. 2002, \apjl, 581, L1

\bibitem[{{Churazov} {et~al.}(2000){Churazov}, {Forman}, {Jones}, \&
  {B{\"o}hringer}}]{churazov00}
{Churazov}, E., {Forman}, W., {Jones}, C., \& {B{\"o}hringer}, H. 2000, \aap,
  356, 788

\bibitem[{{Finoguenov} \& {Jones}(2000)}]{finoguenov00}
{Finoguenov}, A., \& {Jones}, C. 2000, \apj, 539, 603

\bibitem[{{Finoguenov} {et~al.}(2008){Finoguenov}, {Ruszkowski}, {Jones},
  {Brueggen}, {Vikhlinin}, \& {Mandel}}]{finoguenov08}
{Finoguenov}, A., {Ruszkowski}, M., {Jones}, C., {Brueggen}, M., {Vikhlinin},
  A., \& {Mandel}, E. 2008, ArXiv e-prints, 807

\bibitem[{{Forman} {et~al.}(1985){Forman}, {Jones}, \& {Tucker}}]{forman85}
{Forman}, W., {Jones}, C., \& {Tucker}, W. 1985, \apj, 293, 102

\bibitem[{{Forman} {et~al.}(2005){Forman}, {Nulsen}, {Heinz}, {Owen}, {Eilek},
  {Vikhlinin}, {Markevitch}, {Kraft}, {Churazov}, \& {Jones}}]{forman05}
{Forman}, W., {Nulsen}, P., {Heinz}, S., {Owen}, F., {Eilek}, J., {Vikhlinin},
  A., {Markevitch}, M., {Kraft}, R., {Churazov}, E., \& {Jones}, C. 2005, \apj,
  635, 894

\bibitem[{{Hoeft} \& {Br{\"u}ggen}(2004)}]{hoeft04}
{Hoeft}, M., \& {Br{\"u}ggen}, M. 2004, \apj, 617, 896

\bibitem[{{Jetha} {et~al.}(2008){Jetha}, {Hardcastle}, {Babul}, {O'Sullivan},
  {Ponman}, {Raychaudhury}, \& {Vrtilek}}]{jetha08}
{Jetha}, N.~N., {Hardcastle}, M.~J., {Babul}, A., {O'Sullivan}, E., {Ponman},
  T.~J., {Raychaudhury}, S., \& {Vrtilek}, J. 2008, \mnras, 384, 1344

\bibitem[{{Jones} {et~al.}(2002){Jones}, {Forman}, {Vikhlinin}, {Markevitch},
  {David}, {Warmflash}, {Murray}, \& {Nulsen}}]{jones02}
{Jones}, C., {Forman}, W., {Vikhlinin}, A., {Markevitch}, M., {David}, L.,
  {Warmflash}, A., {Murray}, S., \& {Nulsen}, P.~E.~J. 2002, \apjl, 567, L115

\bibitem[{{Kraft} {et~al.}(2004){Kraft}, {Forman}, {Churazov}, {Laslo},
  {Jones}, {Markevitch}, {Murray}, \& {Vikhlinin}}]{kraft04}
{Kraft}, R.~P., {Forman}, W.~R., {Churazov}, E., {Laslo}, N., {Jones}, C.,
  {Markevitch}, M., {Murray}, S.~S., \& {Vikhlinin}, A. 2004, \apj, 601, 221

\bibitem[{{Loewenstein} {et~al.}(2001){Loewenstein}, {Mushotzky}, {Angelini},
  {Arnaud}, \& {Quataert}}]{loewenstein01}
{Loewenstein}, M., {Mushotzky}, R.~F., {Angelini}, L., {Arnaud}, K.~A., \&
  {Quataert}, E. 2001, \apjl, 555, L21

\bibitem[{{Machacek} {et~al.}(2006){Machacek}, {Nulsen}, {Jones}, \&
  {Forman}}]{machachek06}
{Machacek}, M., {Nulsen}, P.~E.~J., {Jones}, C., \& {Forman}, W.~R. 2006, \apj,
  648, 947

\bibitem[{{Markevitch} {et~al.}(2003){Markevitch}, {Mazzotta}, {Vikhlinin},
  {Burke}, {Butt}, {David}, {Donnelly}, {Forman}, {Harris}, {Kim}, {Virani}, \&
  {Vrtilek}}]{markevitch03}
{Markevitch}, M., {Mazzotta}, P., {Vikhlinin}, A., {Burke}, D., {Butt}, Y.,
  {David}, L., {Donnelly}, H., {Forman}, W.~R., {Harris}, D., {Kim}, D.-W.,
  {Virani}, S., \& {Vrtilek}, J. 2003, \apjl, 586, L19

\bibitem[{{Matsushita} {et~al.}(1997){Matsushita}, {Makishima}, {Rokutanda},
  {Yamasaki}, \& {Ohashi}}]{matsushita97}
{Matsushita}, K., {Makishima}, K., {Rokutanda}, E., {Yamasaki}, N.~Y., \&
  {Ohashi}, T. 1997, \apjl, 488, L125+

\bibitem[{{McNamara} \& {Nulsen}(2007)}]{mcnamara07}
{McNamara}, B.~R., \& {Nulsen}, P.~E.~J. 2007, \araa, 45, 117

\bibitem[{{Merritt} \& {Ferrarese}(2001)}]{merritt01}
{Merritt}, D., \& {Ferrarese}, L. 2001, \apj, 547, 140

\bibitem[{{Ohto} {et~al.}(2003){Ohto}, {Kawano}, \& {Fukazawa}}]{ohto03}
{Ohto}, A., {Kawano}, N., \& {Fukazawa}, Y. 2003, \pasj, 55, 819

\bibitem[{{O'Sullivan} {et~al.}(2003){O'Sullivan}, {Ponman}, \&
  {Collins}}]{osullivan03}
{O'Sullivan}, E., {Ponman}, T.~J., \& {Collins}, R.~S. 2003, \mnras, 340, 1375

\bibitem[{{O'Sullivan} {et~al.}(2005){O'Sullivan}, {Vrtilek}, \&
  {Kempner}}]{osullivan05}
{O'Sullivan}, E., {Vrtilek}, J.~M., \& {Kempner}, J.~C. 2005, \apjl, 624, L77

\bibitem[{{Smith} {et~al.}(2001){Smith}, {Brickhouse}, {Liedahl}, \&
  {Raymond}}]{smith01}
{Smith}, R.~K., {Brickhouse}, N.~S., {Liedahl}, D.~A., \& {Raymond}, J.~C.
  2001, \apjl, 556, L91

\bibitem[{{Stark} {et~al.}(1992){Stark}, {Gammie}, {Wilson}, {Bally}, {Linke},
  {Heiles}, \& {Hurwitz}}]{stark92}
{Stark}, A.~A., {Gammie}, C.~F., {Wilson}, R.~W., {Bally}, J., {Linke}, R.~A.,
  {Heiles}, C., \& {Hurwitz}, M. 1992, \apjs, 79, 77

\bibitem[{{Tonry} {et~al.}(2001){Tonry}, {Dressler}, {Blakeslee}, {Ajhar},
  {Fletcher}, {Luppino}, {Metzger}, \& {Moore}}]{tonry01}
{Tonry}, J.~L., {Dressler}, A., {Blakeslee}, J.~P., {Ajhar}, E.~A., {Fletcher},
  A.~B., {Luppino}, G.~A., {Metzger}, M.~R., \& {Moore}, C.~B. 2001, \apj, 546,
  681

\bibitem[{{Trinchieri} {et~al.}(1994){Trinchieri}, {Kim}, {Fabbiano}, \&
  {Canizares}}]{trinchieri94}
{Trinchieri}, G., {Kim}, D.-W., {Fabbiano}, G., \& {Canizares}, C.~R.~C. 1994,
  \apj, 428, 555

\bibitem[{{Vikhlinin}(2006)}]{vikhlinin06}
{Vikhlinin}, A. 2006, \apj, 640, 710

\bibitem[{{Vikhlinin} {et~al.}(2005){Vikhlinin}, {Markevitch}, {Murray},
  {Jones}, {Forman}, \& {Van Speybroeck}}]{vikhlinin05}
{Vikhlinin}, A., {Markevitch}, M., {Murray}, S.~S., {Jones}, C., {Forman}, W.,
  \& {Van Speybroeck}, L. 2005, \apj, 628, 655

\end{thebibliography}

%\end{document}

%\clearpage

%\clearpage

%\begin{figure}
%\epsscale{.80}
%\plotone{dspec_wedge_185_235.ps}
%\caption{Projected profiles (blue) and deprojected profiles (red) for the X-ray gas observed in NGC~4636 
%in the $185^\circ$-$235^\circ$ sector. The orientation of the sector 
%was chosen to avoid any the SW and NE cavities.{\it Top:} 
%Temperature profile. {\it Bottom:} Density profile. \label{deprojSE}}
%\end{figure}

%\clearpage

%\clearpage

%\clearpage

\end{document}